\documentclass[12pt]{iopart}
\usepackage{graphicx}
\usepackage{amssymb}
\usepackage[usenames]{color} 
\begin{document}

\topical[Neutrino-driven wind simulations and nucleosynthesis of heavy
elements]{Neutrino-driven wind simulations and nucleosynthesis of
  heavy elements}

\author{A.~Arcones} 

\address{Institut f\"ur Kernphysik, Technische Universit\"at
  Darmstadt, Schlossgartenstra{\ss}e 2, D-64289 Darmstadt, Germany}

\address{GSI Helmholtzzentrum f\"ur Schwerionenforschung GmbH,
  Planckstr. 1 D-64291 Darmstadt, Germany}

\ead{almudena.arcones@physik.tu-darmstadt.de}

\author{F.-K.~Thielemann}
\address{Department of Physics, University of Basel,
  Klingelbergstra{\ss}e 82, 4056, Basel, Switzerland}

\begin{abstract}
  Neutrino-driven winds, which follow core-collapse supernova
  explosions, present a fascinating nuclear astrophysics problem that
  requires understanding advanced astrophysics simulations, the
  properties of matter and neutrino interactions under extreme
  conditions, the structure and reactions of exotic nuclei, and
  comparisons against forefront astronomical observations. The
  neutrino-driven wind has attracted vast attention over the last 20
  years as it was suggested to be a candidate for the astrophysics
  site where half of the heavy elements are produced via the
  r-process. In this review, we summarize our present understanding of
  neutrino-driven winds from the dynamical and nucleosynthesis
  perspectives.  Rapid progress has been made during recent years in
  understanding the wind with improved simulations and better micro
  physics. The current status of the fields is that hydrodynamical
  simulations do not reach the extreme conditions necessary for the
  r-process and the proton or neutron richness of the wind remains to
  be investigated in more detail. However, nucleosynthesis studies and
  observations point already to neutrino-driven winds to explain the
  origin of lighter heavy elements, such as Sr, Y, Zr.

\end{abstract}

\submitto{\JPG}
\maketitle


\section{Introduction}
\label{sec:intro}
Core-collapse supernovae contribute to the chemical enrichment of the
interstellar medium (and thus to the next generation of stars) in two
ways: they eject elements that were synthesized during the life of
stars (e.g., helium, nitrogen, oxygen and carbon) and they produce new
and heavier elements during the explosion.  Although the link between
core-collapse supernovae and the origin of heavy elements was done
more than fifty years ago \cite{Burbidge.Burbidge.ea:1957,
  Cameron:1957}, there have been many new and exciting developments in
the field of supernovae nucleosynthesis. Here we want to summarize the
nucleosynthesis in the neutrino-powered ejecta known as
neutrino-driven wind of the newly born proto-neutron star.

The lightest elements were produced in the big bang, followed by the
formation of the first stars and by the production of heavier elements
in subsequent nuclear burning phases in their interiors. The death of
the first stars enriches the interstellar medium, and the next
generations of stars reprocess the previously expelled material. This
process continues in recurrent star formation, stellar evolution, and
stellar death stages (galactic chemical evolution). Although this
synthesis is believed to be generally understood, key questions in the
physics remain. For example, how do massive stars die in core-collapse
supernova explosions? How are half of the elements heavier than iron
produced? 

Core-collapse supernovae mark the end of the life of stars with at
least eight times the mass of our sun, leading to the birth of neutron
stars and stellar-mass black holes.  After millions of years of
hydrostatic burning, no further energy can be gained from fusion and
the iron core collapses when it reaches the maximum mass stabilized by
the pressure of a degenerate electron gas. The collapse suddenly
stops, when the core is compressed to nuclear densities by gravity,
and the inner core bounces back, forming a shock wave. This supernova
shock loses energy by photo-dissociation of iron-group nuclei in the
material encountered by the passing shock wave. This leads to a
stalled shock, and it remains an open question how it re-accelerates
and produces a successful explosion (see Refs.~\cite{Janka.etal:2007,
  Janka:2012}). The best studied and still most promising mechanism to
re-accelerate the shock is due neutrinos because they can transport
the energy from the hot proto-neutron star to the shock.

After the successful launch of the supernova explosion, the
proto-neutron star in the center cools by emitting neutrinos. The
energy deposited by these neutrinos via capture and scattering events
powers a baryonic outflow that expands with supersonic velocities and
is known as the neutrino-driven wind. This neutrino-driven wind is a
promising site for different nucleosynthesis processes and was
proposed as the main host for the r-process.  The general conditions
required for the r-process can be studied using
analytic~\cite{Qian.Woosley:1996} and
steady-state~\cite{Otsuki.Tagoshi.ea:2000,
  Thompson.Burrows.Meyer:2001} models of neutrino-driven winds. These
have established high entropy, fast expansion, and low electron
fraction ($Y_e \approx 0.4$) as necessary to obtain a high ratio of
neutrons to so-called seed nuclei in the iron group or beyond
(neutron-to-seed ratio) which act as seeds for rapid neutron capture
to form the heaviest elements.  Parametric models show that a strong
r-process can occur~\cite{hoffman.woosley.qian:1997,
  Freiburghaus.Rembges.ea:1999, Farouqi.etal:2010}, but only for
conditions that are not reached in current long-time hydrodynamic
simulations~\cite{arcones.janka.scheck:2007,
  Fischer.etal:2010,Huedepohl.etal:2010,Roberts.etal:2010}.

At present it is unclear if the conditions in neutrino-driven winds
are extreme enough for a successful r-process up to uranium, but it is
certain that core-collapse supernovae are fascinating hosts where
various nucleosynthesis processes produce neutron-rich and
neutron-deficient nuclei. As suggested in
Ref.~\cite{Qian.Wasserburg:2007}, neutrino-driven winds can also be
the site where lighter heavy elements (e.g., Sr, Y, Zr) are produced
by the weak r-process or by the $\nu$p-process. These can account for
the suggested lighter element primary process (LEPP)
contribution~\cite{Travaglio.Gallino.ea:2004}, responsible for the
abundances of these elements at low metallicities in very old stars
where the s-process is not active yet.

\subsection{Observational constraints}
\label{sec:obs}

The fingerprints of supernova nucleosynthesis are observed in our
solar system. The solar photosphere and meteorites reflect the
chemical signature of the gas cloud where the Sun formed. These
abundances show the combined results of different nucleosynthesis
contributions and the imprints of nuclear physics. The fast decrease
of the abundances towards the iron peak is due to the increasing
Coulomb barrier for charged particle fusion reactions (with increasing
proton number) that hinders the production of heavier elements. Then
neutron capture takes over as main mechanism for forming heavier than
iron-group nuclei, and one distinguishes between the s-process and
r-process~\cite{Burbidge.Burbidge.ea:1957}, which yield double peaks
in the abundances corresponding to neutron magic numbers, $N=$~50, 82,
and 126.

The oldest stars observed, known as ultra metal-poor (UMP) stars, show
lines of heavy elements in their spectra indicating that these
elements were already expelled in very early r-process
events~\cite{Sneden.etal:2008}.  These UMP stars are very rare,
however their detection is increasing successful with large-scale
surveys and new telescopes ~\cite{HERES_II, SDSS:2011} that are
providing new insights in the origin of elements. When abundances of
UMP stars are compared to the scaled solar system abundances, there
are two clear underlying trends: 1) For elements between barium and
lead ($56 < Z< 82$) the relative abundances are the same in UMP stars
and in the r-process component of the solar system. This indicates
that these elements are always produced in the same way by a
\emph{robust r-process}. This robustness can be due to the
astrophysical scenario in combination with nuclear physics aspects. 2)
The scatter for elements between strontium and silver ($38<Z<47$)
indicates an additional contribution for the production of those
\emph{lighter heavy elements}. This contribution may be associated
with a weak r-process (see Ref.~\cite{Sneden.etal:2008} and references
therein) involving charged-particle
reactions~\cite{Qian.Wasserburg:2007}.

The strong scatter of r-process contributions in comparison to regular
supernova products like iron, measured e.g., by the Eu/Fe-ratio, seems
to give an indication that strong r-process events are rare but
efficient when they occur~\cite{Cowan.Thielemann:2004}.
 
\subsection{Structure of the review and definitions}
\label{sec:def}
This review is divided into two parts. First, we discuss the main
features, parameters and nucleosynthesis of the neutrino-driven
outflows (Sect.~\ref{sec:winds}). The wind dynamics, which is key for
determining the entropy and expansion time scale, are described in
Sect.~\ref{sec:wind_dyn}, while weak reactions and electron fraction
uncertainties are introduced in Sect.~\ref{sec:weak_int}. The role of
these three wind parameters (entropy, expansion time scale, and
electron fraction) on the nucleosynthesis is explained in
Sect.~\ref{sec:nuc_par}. The possible additional ingredients that have
been investigated are presented in Sect.~\ref{sec:add_ingredients}.
In the second part (Sect.~\ref{sec:nuc_wind}), the different
nucleosynthesis processes and their status based on recent supernova
simulations are discussed. The developments of neutrino-driven wind
models for studying the r-process are summarized in
Sect.~\ref{sec:rprocess}. The origin of lighter heavy elements
(Sect.~\ref{sec:lepp}) can be explained by a weak r-process
(Sect.~\ref{sec:weak_rprocess}) or by the $\nu p$-process
(Sect.~\ref{sec:nup-process}). Summary and outlook are in
Sect.~\ref{sed:sum}.

Before immersing us into the neutrino-driven wind, lets summarize a
few useful concepts that will be used along the review.

Matter consisting of a composition of ionized nuclei can be described
by the corresponding mass fractions $X_i$ summing up to unity ($\sum_i
X_i = 1$). A quantity which is not weighted with the total mass but
rather proportional to the number density of a nucleus is the
abundance $Y_i=X_i/A_i$.  If the mass density of a nucleus $\rho X_i$
is divided by the mass of a nucleus $A_im_u$, one obtains directly the
number density $n_i=\rho X_i/(A_im_u)=\rho Y_i/m_u=\rho N_A Y_i$, as
$N_A=1/m_u$ in the appropriate units. The total abundance of electrons
(or also called electron fraction) is identical to the total abundance
of protons (in free protons and nuclei) $Y_e=\sum_iZ_iY_i$ and
equivalent to the total proton-to-nucleon ratio $\sum_iZ_iY_i /
(\sum_i(Z_i+N_i)Y_i) = \sum_iZ_iY_i / ( \sum_i A_i Y_i) = \sum_iZ_iY_i
/ (\sum_iX_i)$. Abundance changes ($\dot{Y}_i$) can be described by
differential equations for each nuclear abundance $Y_i$, related to
decays, fusion reactions and three body reactions (a sequence of two
fusion reactions involving an intermediate particle-unstable nucleus,
i.e. with a vanishing half-life), for details see
Ref.~\cite{Hix.Thielemann:1999}.  Reactions with a (thermal or
non-thermal) distribution of photons/neutrinos can be written like
decay reactions (after integrating over the photon or neutrino energy
spectra) with a decay ``constant'' having a temperature (or more
complex) dependence.

\section{Neutrino-driven outflows: features, parameters and
  nucleosynthesis}
\label{sec:winds}

\subsection{Wind dynamics: entropy and expansion time scale}
\label{sec:wind_dyn}

When a massive star collapses at the end of its live, gravitational
energy is transformed into internal energy. This leads to a hot
proto-neutron star with initial temperatures of $kT \approx GM m_n/R
\approx 30-50$~MeV. At this high temperature, neutrinos dominate the
cooling and carry away the gravitational binding energy. Inside of the
newly born neutron star, densities are very high and neutrinos are
thus trapped. Neutrinos diffuse from the interior of the neutron star
and can freely stream when their mean free path becomes comparable to
the neutron star radius. The region from where neutrinos escape, the
neutrinosphere, is different for every flavor and energy as
neutrino-matter interactions are strongly energy dependent. The
neutrinosphere region is located at the outer layers of the
proto-neutron star where density and temperature present steep
gradients. Some of the escaping neutrinos deposit energy in the matter
of this region mainly via charged-particle reactions:
\begin{eqnarray}
  \nu_e + n &\rightarrow &p + e^- \label{eq:nue_n}\\
  \bar{\nu}_e + p &\rightarrow &n + e^+ \, .\label{eq:nuebar_p}
\end{eqnarray}
This injection of energy in the outer layers of the proto-neutron star
causes that a significant fraction of mass of the outer neutron star
layers is blown off in a \emph{neutrino-driven wind}.

The neutrino-driven wind sets in after the explosion and lasts for
several seconds and even minutes. This deleptonization eventually
leads to a cold neutron star, transparent to neutrinos, and to a
significant amount of matter being ejected during the first 10--20~s
after the explosion. This initial cooling phase or neutrino-driven
wind has been first described by Duncan et
al. (1986)~\cite{duncan.shapiro.wasserman:1986}, a detailed study of the
later cooling phase can be found in Ref.~\cite{Pons99}.

\begin{figure}[!htb]
  \includegraphics[width=0.55\linewidth]{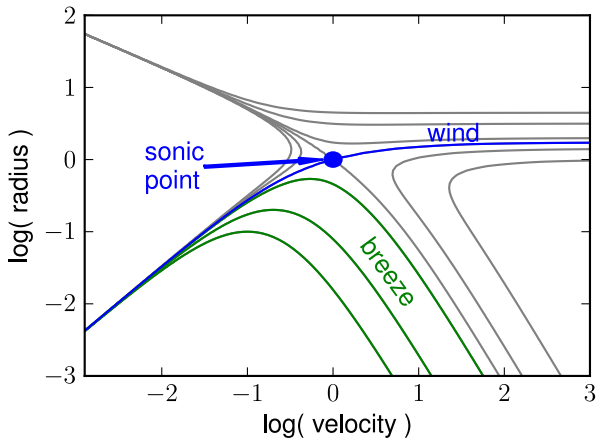}
  \caption{Schematic representation of the solutions of
    Eqs.~(\ref{eq:wind_eq_1})-(\ref{eq:wind_eq_3}). The blue line
    corresponds to the physical supersonic solution that crosses the
    sonic point. Some of the physical subsonic solutions are shown by
    the green lines and labeled as breeze. The grey lines represent
    mathematical solutions without physical meaning.}
  \label{fig:wind_breeze}
\end{figure}

During the Kelvin-Helmholtz cooling time scale (i.e., the time it
takes to radiate away the gravitational energy) the neutrino
luminosities and energies and also the neutron star radius and mass
change only slowly. Therefore, neutrino-driven winds can be assumed to
be steady state outflows \cite{duncan.shapiro.wasserman:1986}
described by the following equations:
\begin{eqnarray}
  \dot{M} & = & 4 \pi r^2 \rho v \label{eq:wind_eq_1}\\
v \frac{dv}{dr} & = & -\frac{1}{\rho} \frac{dP}{dr} - \frac{GM}{r^2} \label{eq:wind_eq_2}\\
\dot{q} & = & v \left( \frac{d \epsilon}{dr} - \frac{P}{\rho^2}\frac{d\rho}{dr} \right)\label{eq:wind_eq_3}
\end{eqnarray}
for the mass, momentum, and energy conservation, respectively. Here
$\dot{M}$ is a constant mass outflow, $\rho$ is the rest mass density,
$v$ is the outflow velocity, $M$ is the mass of the neutron star,
$\dot{q}$ is the energy generation rate produced by neutrinos, and $P$
and $\epsilon$ are pressure and specific energy that account for
non-relativistic nucleons, relativistic electrons and positrons, and
photon radiation \cite{duncan.shapiro.wasserman:1986,
  Qian.Woosley:1996, Janka:2001}. There are two types of solutions for
these equations with physical meaning. 1) For high mass outflows the
velocity reaches the sound speed and shock region and neutron star are
sonically disconnected. This is known as wind and it is schematically
illustrated by the blue line in Fig.~\ref{fig:wind_breeze}. 2) When
the velocity does not attain the critical sound speed, we talk about a
breeze solution (green lines, Fig.~\ref{fig:wind_breeze}). Notice that
there is only one physical supersonic or critical solution while one
can find several subsonic or breeze solutions.

Steady state and analytic models have been developed to better
understand neutrino-driven winds and their potential as r-process
site, after the promising results of Woosley et al. (1994)
\cite{Woosley.etal:1994}. Qian \& Woosley (1996)
\cite{Qian.Woosley:1996} identified by means of an analytic model the
wind parameters that are key for the nucleosynthesis of heavy
elements.  They assumed the wind to be a steady-state spherical
outflow with boundaries at the neutrinosphere and supernova
shock. Seconds after the explosion the shock is at large radii and
there are only small and slow variations in the global neutrino
characteristics and the properties of the neutron star. These support
the steady-state assumption and permit to describe the wind
independently of details how the shock gets launched. The impact of
neutron star mass ($M_{ns}$) and radius ($R_{ns}$), and the neutrino
luminosity ($L_{\nu}$) and energy spectra ($\epsilon_{\nu}$) on the
wind parameters can be understood with the following
relations~\cite{Qian.Woosley:1996}:
\begin{eqnarray}
  \dot{M} &\propto& L_{\nu}^{5/3}\, \epsilon_{\nu}^{10/3} \, R_{ns}^{5/3} \, M_{ns}^{-2} \, , \label{eq:mdot_qw}\\
  s &\propto& L_{\nu}^{-1/6}\, \epsilon_{\nu}^{-1/3} \, R_{ns}^{-2/3} \, M_{ns} \, , \label{eq:swind_qw}\\
  \tau &\propto& L_{\nu}^{-1} \, \epsilon_{\nu}^{-2} \, R_{ns} \, M_{ns} \, . \label{eq:tau_qw}
\end{eqnarray}
The mass outflow ($\dot{M}$) gives the amount of ejected mass and thus
the contribution of neutrino-driven winds to the enrichment of heavy
elements in the universe, provided the conditions in these mass zones
support their nucleosynthesis. The wind entropy ($s$) and the
expansion time scale ($\tau$) are critical parameters to determine the
composition of the ejecta.

The reactions responsible for the synthesis of heavy nuclei start at
$^{12}$C which is produced from $^{4}$He via three body reactions (see
Sect.~\ref{sec:nuc_par}). The efficiency of these reactions depends on
the range of temperatures and densities. In a radiation-dominated
environment the entropy is given by a relation including temperature
and density: $S_w \propto T^3/\rho$. This is proportional to the
photon-to-baryon ratio \cite{Meyer.Mathews.ea:1992, Meyer:1993}. At
high entropies the temperature is also high for densities which permit
three body reactions. High temperature causes photo-dissociation and
prevents the formation of seed elements (Sect.~\ref{sec:def}).  The
expansion time scale also controls the efficiency of the three body
reactions. If the expansion is very fast, alpha particles do not have
sufficient time to combine into seed nuclei. Therefore, entropy and
expansion time scale are key to determine how alpha particles
contribute to form seed nuclei.

\subsection{Weak interactions: electron fraction}
\label{sec:weak_int}

An additional wind parameter is the electron fraction which does not
depend on the total neutrino luminosities and energies but on the
relative contributions of neutrinos and antineutrinos to these
quantities. In a neutrino-driven wind the two forward reactions in
Eqs.~(\ref{eq:nue_n}) and (\ref{eq:nuebar_p}) dominate and cause,
e.g., the proton abundance time variation to be
\begin{equation}
  \dot{Y}_p = -\lambda_{\bar{\nu}_e,p}Y_p + \lambda_{\nu_e,n} Y_n \, .
\end{equation} 
In case of a weak equilibrium, i.e., an equilibrium of production of
neutrons and protons by these weak interaction reactions, we have
$\dot{Y}_p=\dot{Y}_n=0$. This gives a relation between $Y_p$ and $Y_n$
to be $Y_n/Y_p=\lambda_{\bar{\nu}_e,p} / \lambda_{\nu_e,n}$. On the
other hand, the electron fraction $Y_e$ (the electron to nucleon
ratio, being in charge equilibrium equal to the proton to nucleon
ratio),
\begin{equation}
  Y_e = \frac{Y_p}{Y_p+Y_n} = \frac{1}{1+\frac{Y_n}{Y_p}}
\end{equation}
can be expressed as
\begin{equation}
  Y_e = \frac{1}{1+\frac{\lambda_{\bar{\nu}_e, p}}{\lambda_{\nu_e, n}}}\, .
\end{equation}
Thus, for the the wind one can assume that the electron fraction is
given by \cite{Qian.Woosley:1996}:
\begin{equation}
  Y_e \approx
  \left[ 1+ 
    \frac{
      L_{\bar{\nu}_e} (\epsilon_{\bar{\nu}_e} -2\Delta +1.2\Delta^{2}/ \epsilon_{\bar{\nu}_e})
    }
    {L_{\nu_e} (\epsilon_{\nu_e} +2\Delta +1.2\Delta^{2} /  \epsilon_{\nu_e})
    }
  \right]^{-1}
  \label{eq:ye_qw}
\end{equation}
where $L_{\nu_e}$, $\epsilon_{\nu_e}$ and $L_{\bar{\nu}_e}$,
$\epsilon_{\bar{\nu}_e}$ are the electron neutrino and antineutrino
luminosities and mean energies ($\epsilon_{\nu}=\langle \varepsilon^2
\rangle/ \langle \varepsilon \rangle$), respectively. The
neutron-proton mass difference is $\Delta = m_n-m_p=1.293$~MeV. In
order to have a neutron-rich wind, i.e., $Y_e<0.5$, the neutrino and
antineutrino energies have to approximately fulfill
$\epsilon_{\bar{\nu}_e} - \epsilon_{\nu_e} \gtrsim 4 \Delta \approx
5$~MeV. Indeed, this is not found in recent spherically symmetric
simulations \cite{Huedepohl.etal:2010, Fischer.etal:2010} where the
neutrino outflow stays proton-rich. Under such conditions the rapid
neutron capture process to form the heaviest nuclei cannot occur,
ruling out the spherically symmetric neutrino-driven winds as its
astrophysical site. However, this conditions become very favourable
for the $\nu p$-process~\cite{Froehlich06}.

\begin{figure}[!htb]
  \includegraphics[width=8cm]{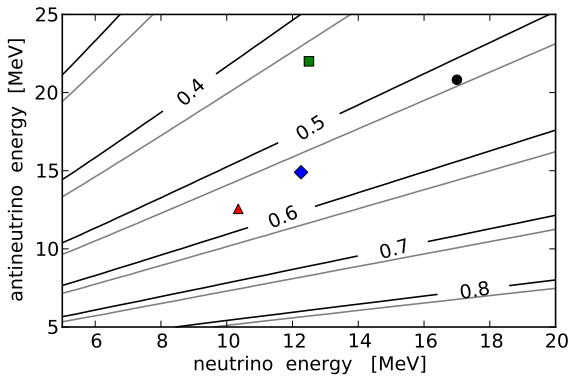}
  \caption{Estimated electron fraction (Eq.~(\ref{eq:ye_qw})) for
    different supernova models. The symbols show the electron neutrino
    and antineutrino energies ($\epsilon_{\nu_e} \approx 4.1
    kT_{\nu_e}$): square for Ref.~\cite{Woosley.etal:1994}, circle for
    model M15-l1-r6 of Ref.~\cite{arcones.janka.scheck:2007}, triangle
    for a 10~$M_{\odot}$ progenitor of Ref.~\cite{Fischer.etal:2010},
    and diamond for Ref.~\cite{Huedepohl.etal:2010}, all at 10~s after
    bounce.  The black contours correspond to
    $L_{\bar{\nu}_e}/L_{\nu_e}=1$ and the grey contours to
    $L_{\bar{\nu}_e}/L_{\nu_e}=1.1$.}
  \label{fig:neut2col}
\end{figure}

Here we would like to shortly summarize the historical evolution of
research efforts for the wind electron fraction. Improvements in the
neutrino reactions have resulted in a reduction of the electron
antineutrino energy from the first simulations by Wilson to the modern
ones \cite{Huedepohl.etal:2010, Fischer.etal:2010}. Moreover, all
neutrino flavours have very similar energies which leads to
proton-rich conditions since the requirement $\epsilon_{\bar{\nu}_e} -
\epsilon_{\nu_e} \gtrsim 4 \Delta \approx 5$~MeV is not
fulfilled. This is illustrated in the neutrino two-color plot
presented in Fig.~\ref{fig:neut2col}. The question about the apparent
proton-richness in neutrino-driven winds can then be reformulated: why
are neutrino energies similar for all neutrino flavours?

Neutrino energies depend on the temperature of the medium in the
region where neutrinos decouple from matter. This region is known as
the neutrinosphere and its location is different for each neutrino
flavor and energy.  Inside of the neutrinosphere, neutrinos are in
thermal and chemical equilibrium mainly through charged-current
reactions (Eqs.~(\ref{eq:nue_n})-(\ref{eq:nuebar_p})).  Outside the
neutrinosphere, neutrinos escape and their temperature stays almost
constant and approximately equal to the temperature at their
neutrinosphere.  Because neutrons are more abundant than protons in
the neutron star, electron neutrinos continue interacting up to larger
radii and thus to lower temperatures than antineutrinos
($\varepsilon_{\bar{\nu}_e} \gtrsim \varepsilon_{\nu_e}$). The muon
and tau (anti)neutrinos interact only via neutral-current reactions
and decouple at smaller radii, therefore their energies are larger.
During the first seconds after the explosion, the proto-neutron star
deleptonizes and the amount of protons in the outer layers decreases.
The electron antineutrino energies are thus expected to be higher than
the electron neutrino energies. However, this simple picture is not
valid and the spectra of electron neutrinos and antineutrinos are
rather similar due to neutral-current reactions that act in a similar
way on all neutrino flavors and become also important as the neutron
star cools ~\cite{Huedepohl.etal:2010,
  arcones.matinezpinedo.etal:2008, Fischer.etal:2012,
  Bacca.etal:2012}.

Moreover, the latest investigations of charged-current neutrino
opacity \cite{MartinezPinedo.etal:2012, Roberts:2012,
  Roberts.Reddy:2012} indicate that these cross sections can
significantly change due to in medium modifications
~\cite{Reddy.etal:1998}. In the early supernova explosion phase, mean
field effects have a minor effect on the neutrino spectra because
neutrinospheres are not at very high densities. In the region where
mean field effects are relevant neutrinos are trapped. In contrast,
during the neutrino-driven wind phase, neutrinospheres are at high
densities ($\rho \approx 10^{12} \mathrm{g} \mathrm{cm}^{-3}$) and
mean field effects are not negligible although they are not included
in current supernova simulations
~\cite{Huedepohl.etal:2010,Fischer.etal:2010}. As suggested in
Refs.~\cite{MartinezPinedo.etal:2012, Roberts:2012,
  Roberts.Reddy:2012}, the consideration of in medium modifications
could change the electron fraction towards more neutron-rich
conditions ($Y_e\approx 0.45$). However, the exact $Y_e$ value depends
on medium correlations and the equation of state, both of which are
still uncertain. In any case, these new results rise an exciting
possibility of linking the wind nucleosynthesis to the behaviour of
matter and neutrinos under extreme conditions in the proto-neutron
star.

In addition, neutrino oscillations (see
Sect.~\ref{sec:add_ingredients} for more details) and the presence of
light clusters such as $^2$H, $^3$H, $^3$He
\cite{arcones.matinezpinedo.etal:2008} also may change the electron
fraction of the wind.

\subsection{Nucleosynthesis and wind parameters}
\label{sec:nuc_par}

A mass elements is ejected from the outer layers of the proto neutron
star due to the energy deposited by neutrino. The nucleosynthesis in
neutrino-driven winds starts at high temperatures and densities which
keep the matter in nuclear statistical equilibrium (NSE), i.e., there
is a balance (i.e., chemical equilibrium) between nuclear reactions
producing seed nuclei $(Z,A)$ and photo-dissociation destroying those
nuclei into nucleons:
\begin{equation}
  (Z,A) \longleftrightarrow (A-Z) n + Z p \, .
\end{equation}
This leads to a relation between the chemical potentials,
$\mu_{(Z,A)}=(A-Z)\mu_{n}+Z\mu_{p}$. Therefore, the composition can be
calculated utilizing Boltzmann distributions for nuclei, neutrons, and
protons. The composition is thus uniquely determined by the
temperature, density, and electron fraction, being environment
parameters:
\begin{equation}
  Y(Z,A)= G_{Z,A} (\rho N_A)^{A-1} \frac{ A^{3/2}}{2^A} 
  \left( \frac{2 \pi \hbar }{m_u kT} \right)^{\frac{3}{2}(A-1)} 
  \exp(B_{Z,A}/kT) Y_n^{A-Z}Y_p^Z
  \label{eq:Y_NSE}
\end{equation}
where $G_{Z,A}$ is the nuclear partition function, $B_{Z,A}$ is the
binding energy, $\hbar$ is the Planck constant, $N_{\mathrm{A}}$ is
the Avogadro number, and $\rho$ is the baryon density. Equation
(\ref{eq:Y_NSE}) indicates that the presence of a nucleus depends
strongly on $kT$, $B(Z,A)/kT$, and $A$. For moderate temperatures
iron-group nuclei are favored in NSE because of their large binding
energies. In contrast, at high temperatures many energetic photons are
available, and the result is a gas consisting mostly of nucleons and
alpha particles in which it is difficult to build up heavy nuclei
because they are quickly photo-dissociated (see
Ref.~\cite{Meyer:1993}). Very high densities would favor in contrast
very heavy nuclei.

The NSE composition at high temperatures is dominated by alpha
particles and nucleons. Only when the temperature drops, the alpha
particles recombine and form seed nuclei at expenses of nucleons whose
abundances drop rapidly. As the expansion continues and matter cools,
slower reactions fall out of equilibrium. At the breakdown of NSE ($T
\sim 8-5 \cdot 10^9$~K) for low densities, as occurring in the
neutrino wind, the composition is dominated by alpha particles and one
talks about alpha-rich freeze out.  Alpha particles combine into seed
nuclei starting with the triple-alpha reaction ($3 \alpha \rightarrow
^{12}\mathrm{C}$). If the amount of free neutrons is not negligible,
$^{12}$C forms also via $^4$He ($\alpha n, \gamma$) $^9$Be ($\alpha,
n$) $^{12}$C~\cite{Woosley.Hoffman:1992}. These reactions depend
strongly on the density and can be hindered if the expansion is very
fast, i.e., if the density does not stay for sufficiently long time in
the range where these three body reactions are effective ($\rho \sim 5
\cdot 10^6 \mathrm{g cm}^{-3}$).  The production of $^{12}$C is
followed by the alpha process \cite{Woosley.Hoffman:1992,
  Witti.etal:1994} which consist of a sequence of alpha captures
(including $(\alpha,\gamma)$, $(\alpha,n)$, and $(\alpha,p)$
reactions) combined with $(n,\gamma)$ and $(p,\gamma)$ reactions
depending on the neutron-richness of the wind.  These charged-particle
reactions (CPR) continue until temperatures are too low to overcome
the Coulomb barrier of nuclei ($T \sim 10^{9}$~K). The composition at
CPR freeze-out consists mainly of alpha particles together with a few
seed nuclei (formed by the alpha-process) and free nucleons. The
proton or neutron richness as well as the ratio of nucleons and seed
nuclei depends on the wind parameters and will determine the
nucleosynthesis processes taking place in neutrino-driven winds. For
example, the r-process can build heavy elements up to Uranium if the
neutron-to-seed ratio ($Y_n/Y_{\mathrm{seed}}$) is sufficiently
high. Assuming the average mass number of seed nuclei is
$\bar{A}_{\mathrm{seed}}$, the neutron-to-seed indicates the heaviest
elements (with mass number $\sim A$) that can be synthesized:
$Y_n/Y_{\mathrm{seed}} +\bar{A}_{\mathrm{seed}} \sim A$. These two key
quantities, $\bar{A}_{\mathrm{seed}}$ and $Y_n/Y_{\mathrm{seed}}$,
depend on the electron fraction, entropy, and expansion time
scale~\cite{Woosley.Hoffman:1992, hoffman.woosley.qian:1997}.
Freiburghaus et al. (1999) \cite{Freiburghaus.Rembges.ea:1999}
analyzed in a site independent, entropy based approach abundance
features and the impact of nuclear physics with especial focus on the
dynamical evolution up to the freeze-out of final neutron captures in
the r-process.

\begin{figure}[!htb]
  \includegraphics[width=6cm]{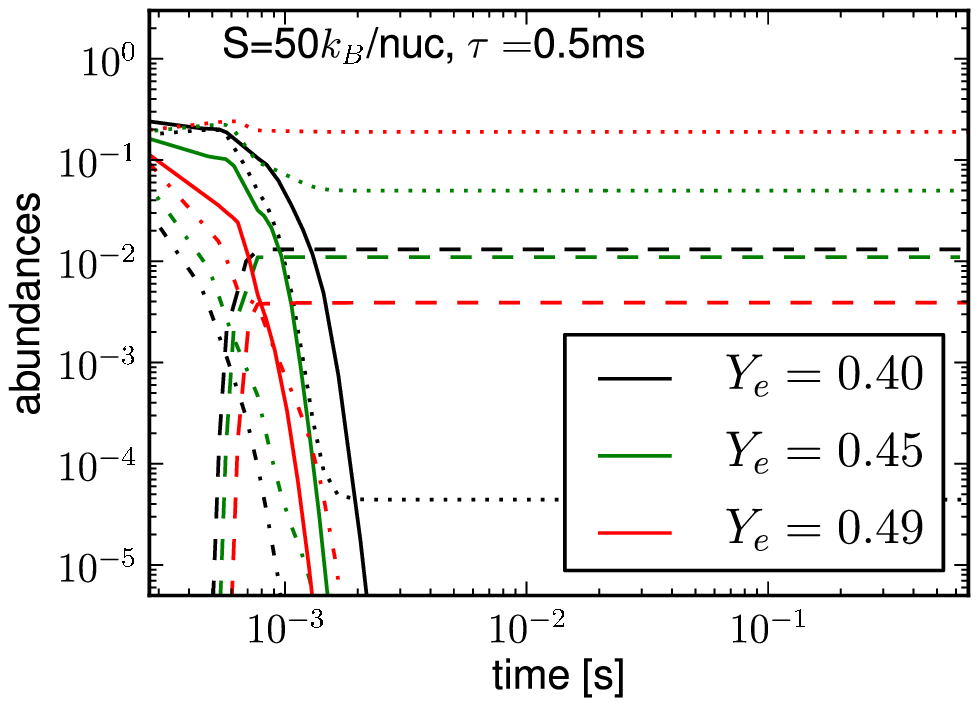}%
  \includegraphics[width=6cm]{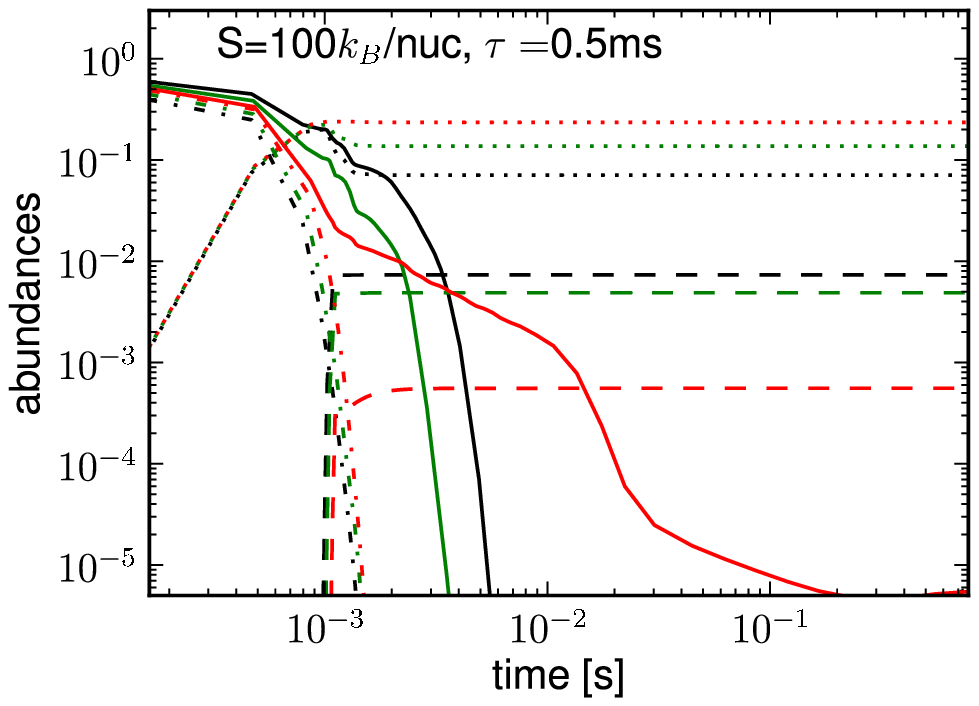}%
  \includegraphics[width=6cm]{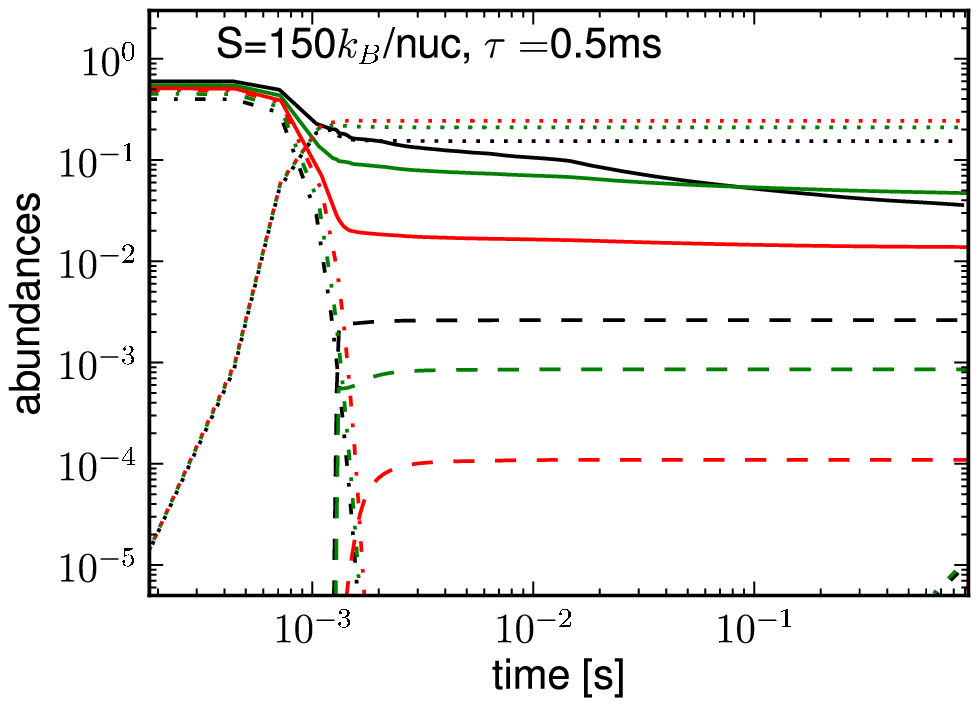}\\
  \includegraphics[width=6cm]{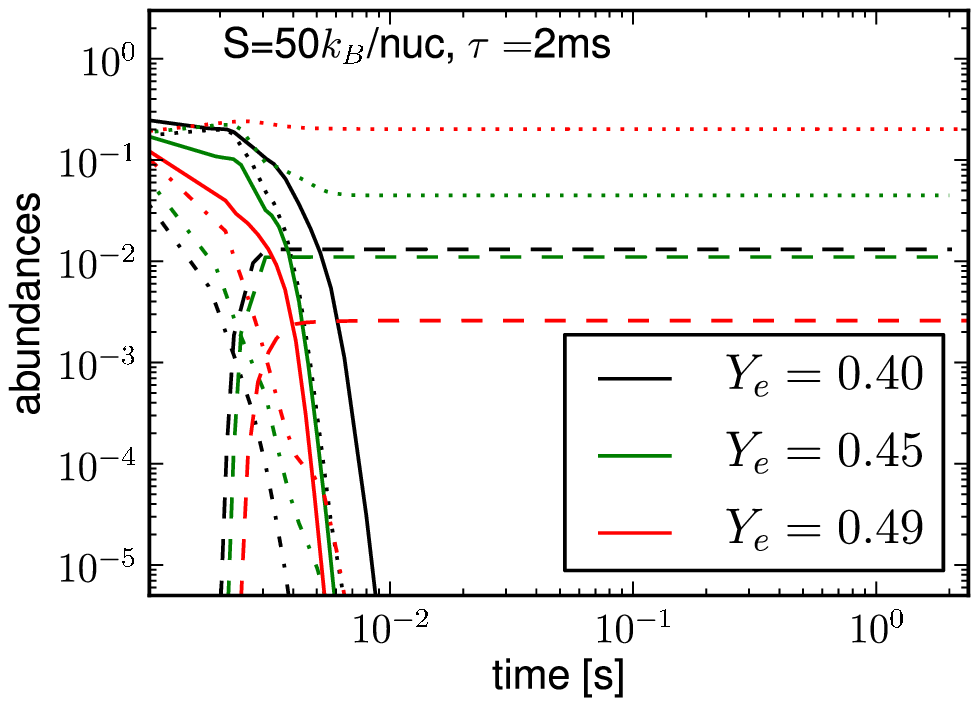}%
  \includegraphics[width=6cm]{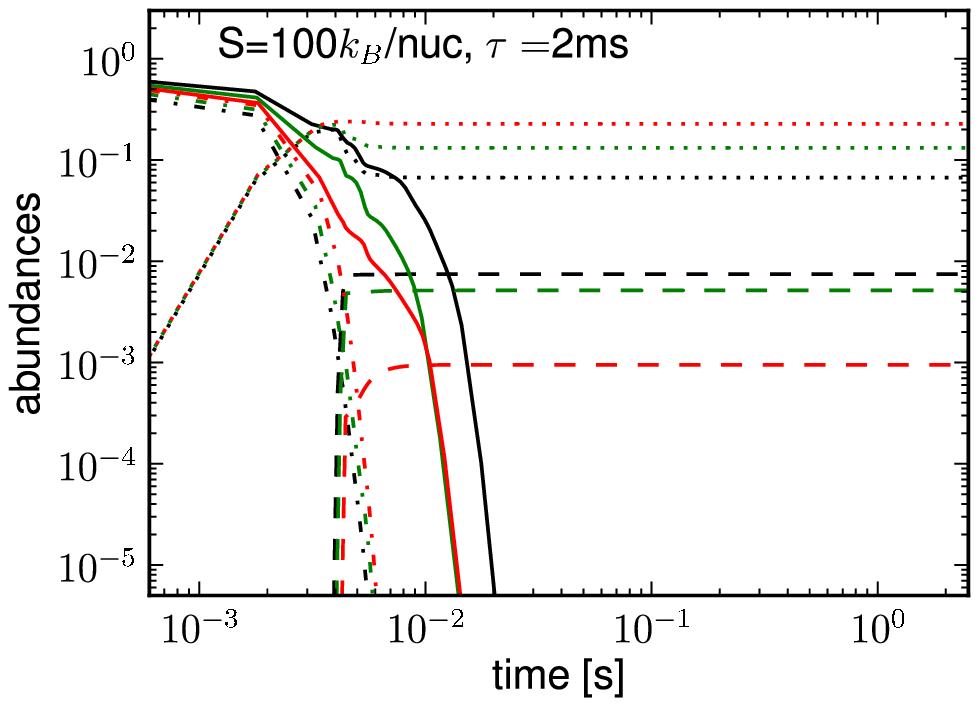}%
  \includegraphics[width=6cm]{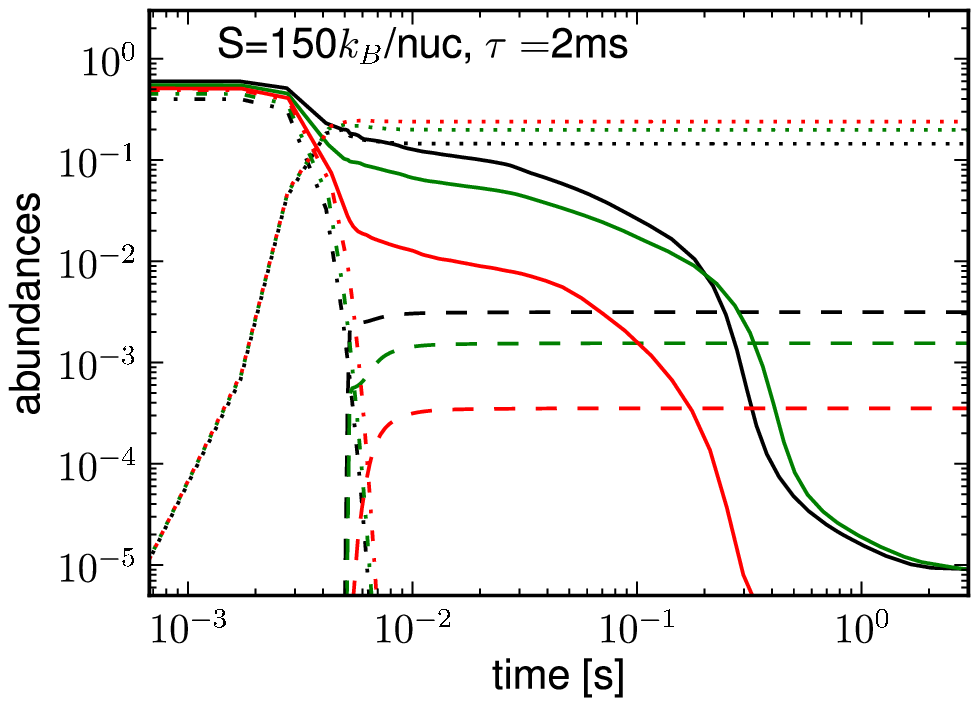}\\
  \includegraphics[width=6cm]{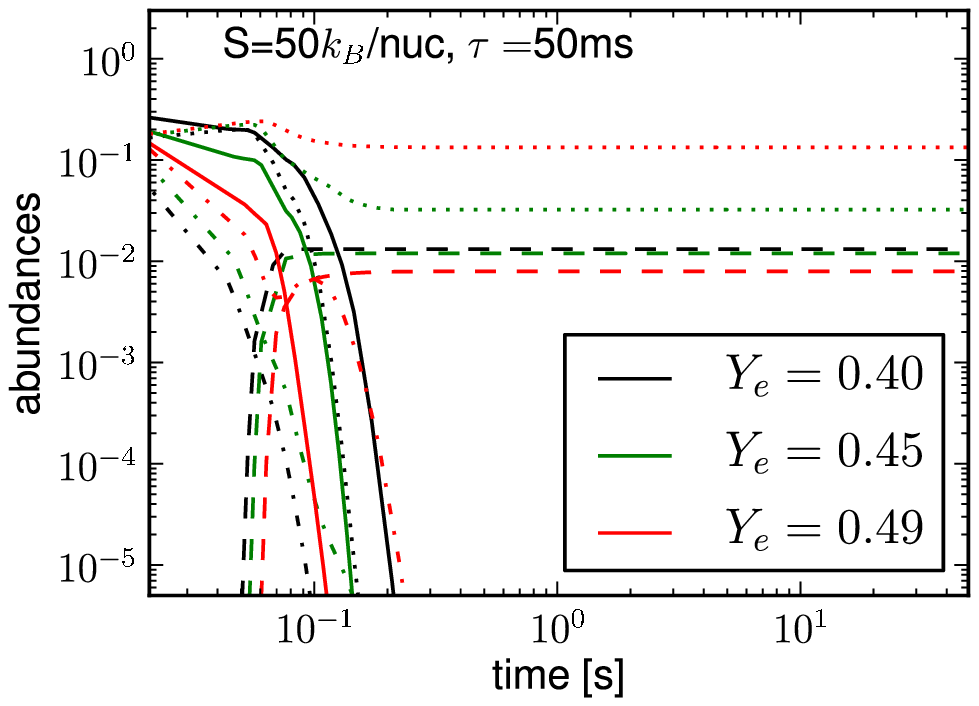}%
  \includegraphics[width=6cm]{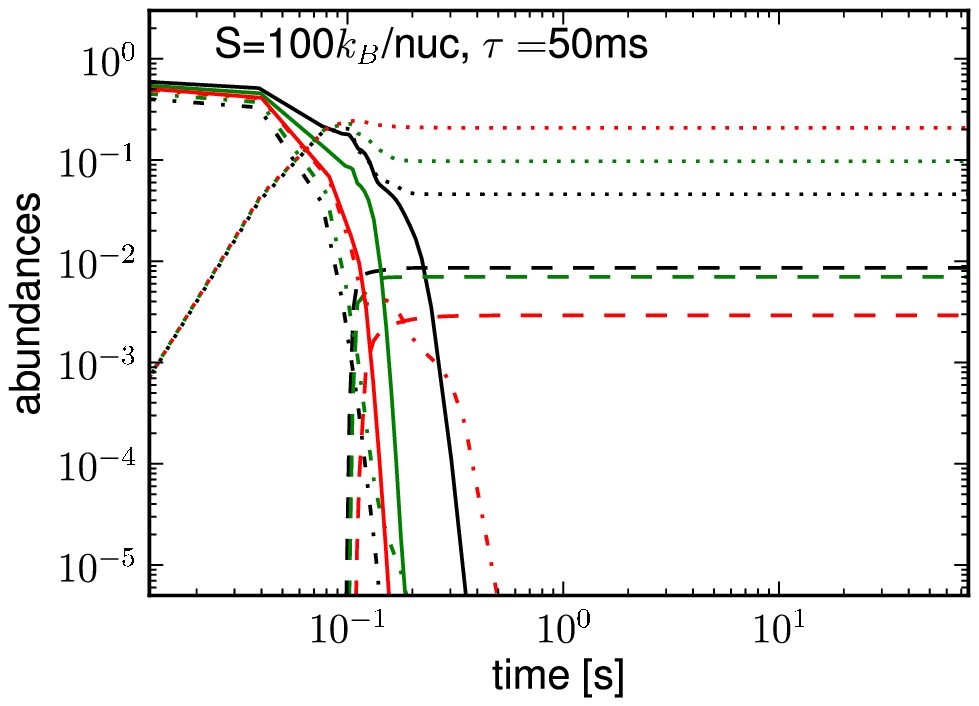}%
  \includegraphics[width=6cm]{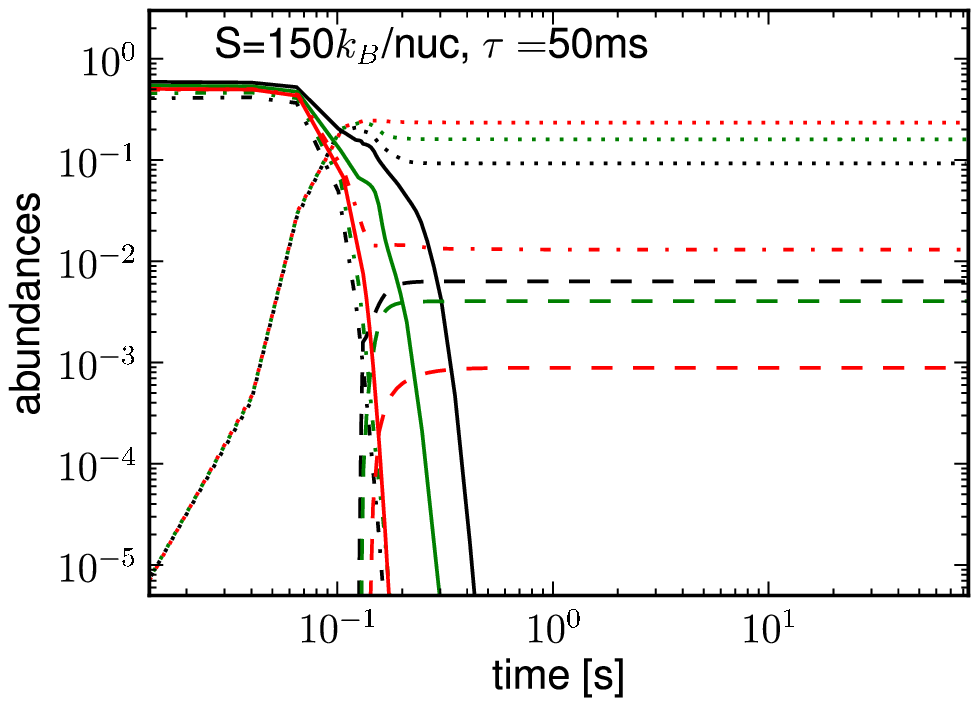}\\
  \caption{Evolution of abundances of neutrons (solid line), protons
    (dashed-dotted line), alpha paticles (dotted line), and heavier
    nuclei (dashed line) for different electron fractions as indicated
    by the colors in the caption. The columns correspond to different
    entropies: $S=50, 100, 150 \, k_{\mathrm{B}}/\mathrm{nuc}$ from
    the left to the right. The three rows are based on expansions with
    varius time scales, as indicated in the panels,
    $\tau=0.5,2,50$~ms. }
  \label{fig:Xevol_nrich}
\end{figure}

In order to understand better the evolution of the composition and its
dependency on wind parameters during and after NSE we present in
Figure~\ref{fig:Xevol_nrich} an overview based on the parametric
trajectories described in~\ref{sec:par_traj}.  Different columns
represent various wind entropies and the rows correspond to various
expansion time scales. For every combination of entropy and time scale
the impact of the electron fraction is also shown by different color
lines. These figures cover the evolution from NSE over
charged-particle freeze-out to the weak or strong r-process. When the
alpha abundances (dotted lines) become constant, the charged-particle
reactions freeze out. At this moment one can get the neutron-to-seed
ratio from the abundances of neutrons (solid lines) and nuclei (dashed
lines). Note that after a few seconds the nuclei (dashed lines) are
not seed nuclei anymore but the product of further neutron capture
reactions.  The impact of the wind parameters can be summarize based
on Figure~\ref{fig:Xevol_nrich}:
\begin{itemize}
\item At high entropies photo-dissociation prevents the formation of
  seed nuclei, therefore the abundance of neutrons stays higher for
  longer times. This behaviour is apparent from the right column of
  the panels with fast expansion. These evolutions will allow for a
  strong r-process. The impact of the wind entropy have been broadly
  discussed in the literature (see e.g., \cite{Meyer:1993,
    Freiburghaus.Rembges.ea:1999, hoffman.woosley.qian:1997}).
\item Fast expansion hinders the three body reactions which mark the
  beginning of the seed nuclei formation. During the production of
  seed nuclei neutrons are used to form alpha particles and heavy
  nuclei. Therefore, a fast expansion increases the neutron-to-seed
  ratio and favours the r-process.
\item The electron fraction directly determined the amount of neutrons
  and protons. Lower $Y_e$ leads to higher $Y_n$ and lower
  $Y_p$. Notice that for $Y_e=0.49$ the behavior of seed nuclei and
  neutrons is not always straight forward, see panel for $S=100
  \, k_{\mathrm{B}}/\mathrm{nuc}$ and $\tau=0.5$~ms. A similar trend was
  seen in Ref.~\cite{hoffman.woosley.qian:1997}.
\end{itemize}

\begin{figure}[!htb]
  \includegraphics[width=6cm]{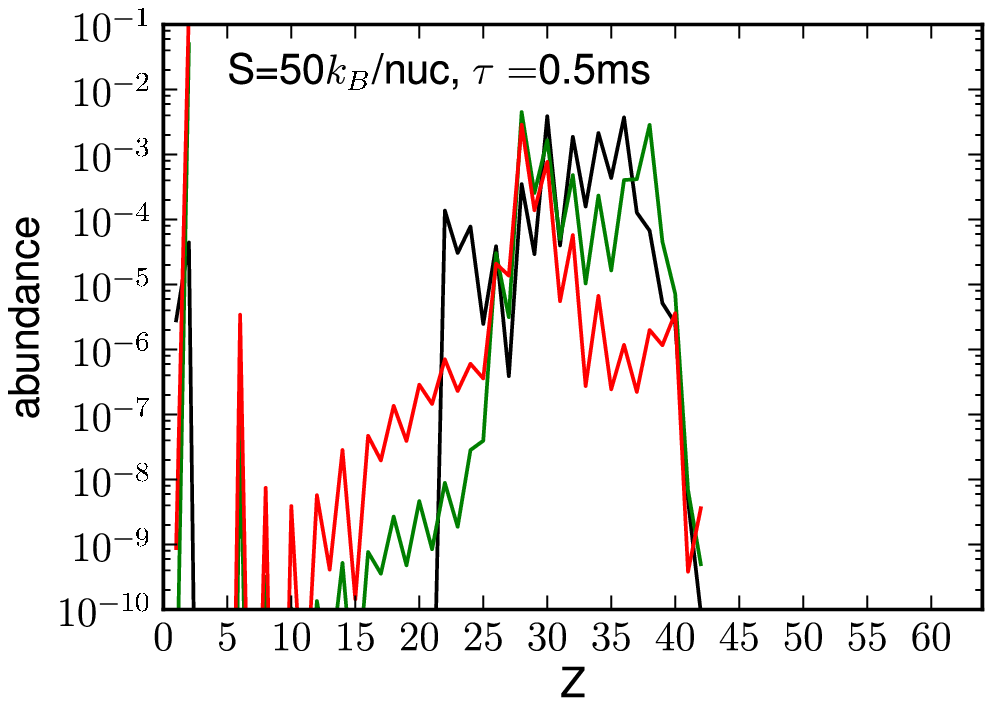}%
  \includegraphics[width=6cm]{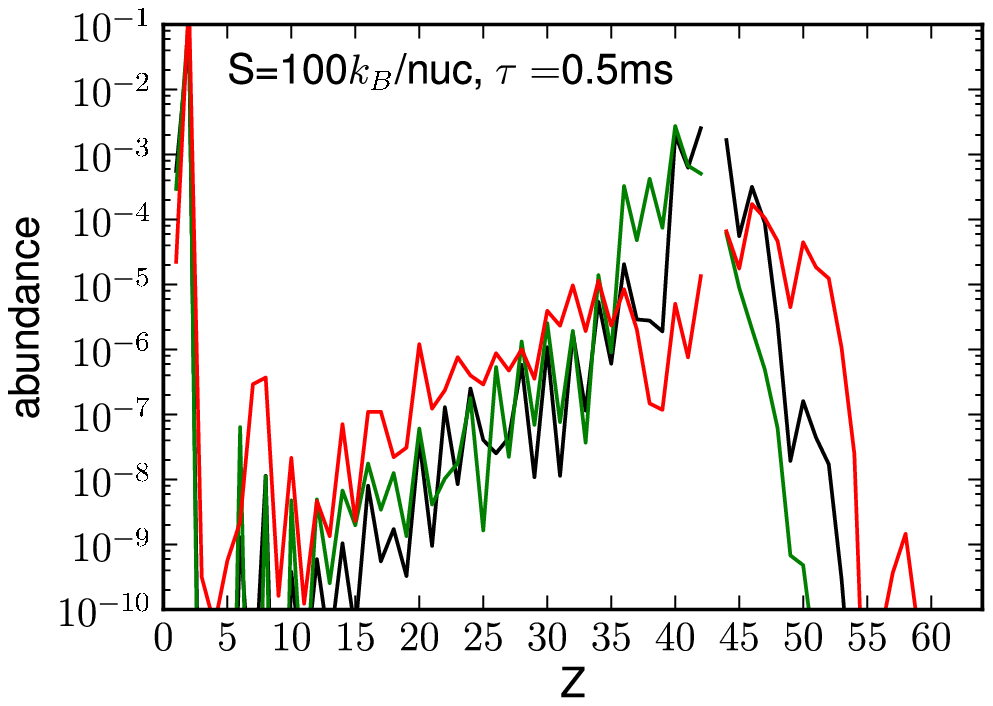}%
  \includegraphics[width=6cm]{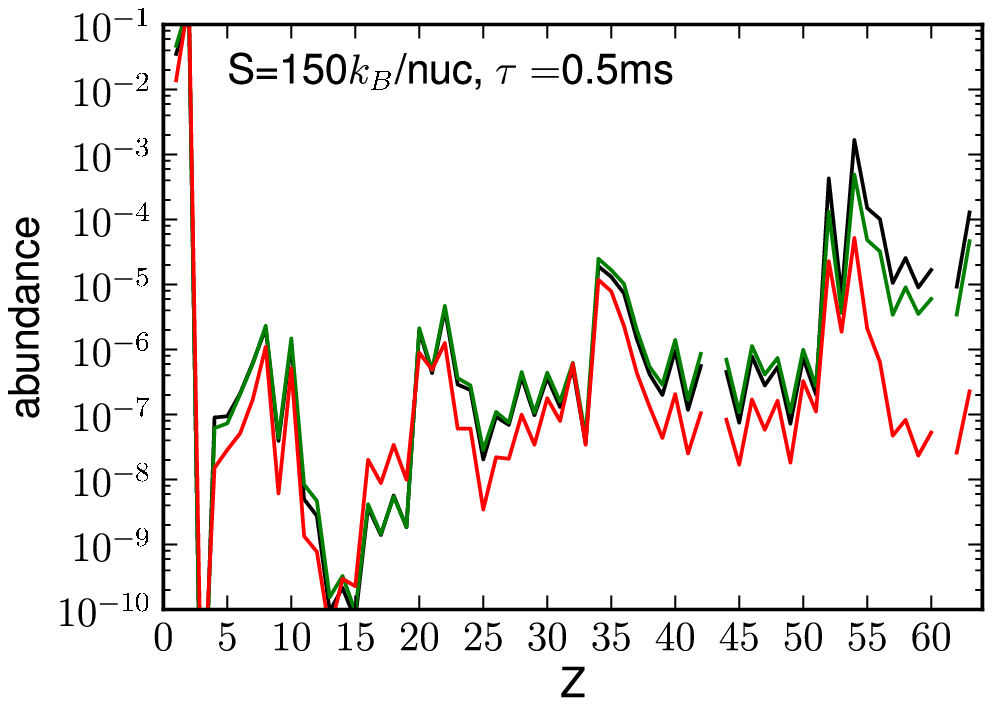}\\
  \includegraphics[width=6cm]{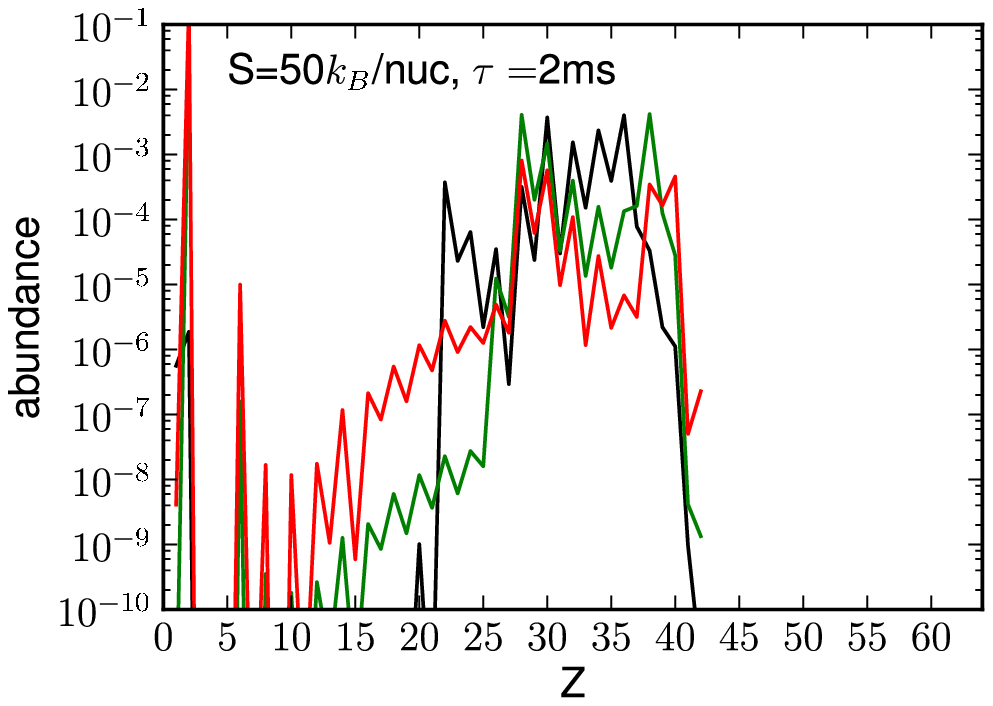}%
  \includegraphics[width=6cm]{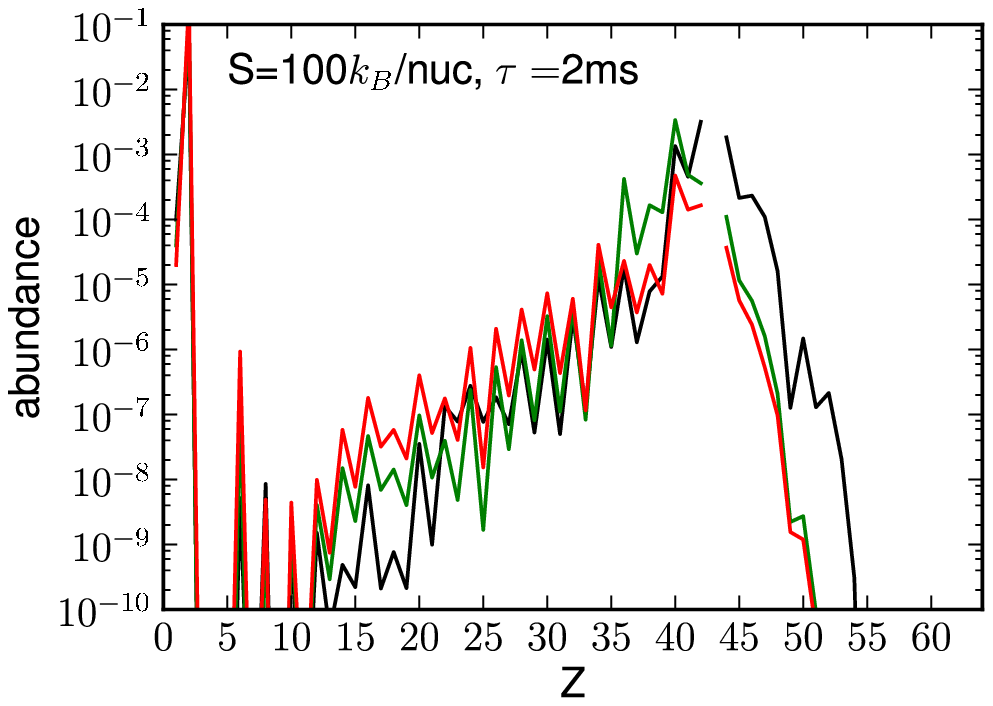}%
  \includegraphics[width=6cm]{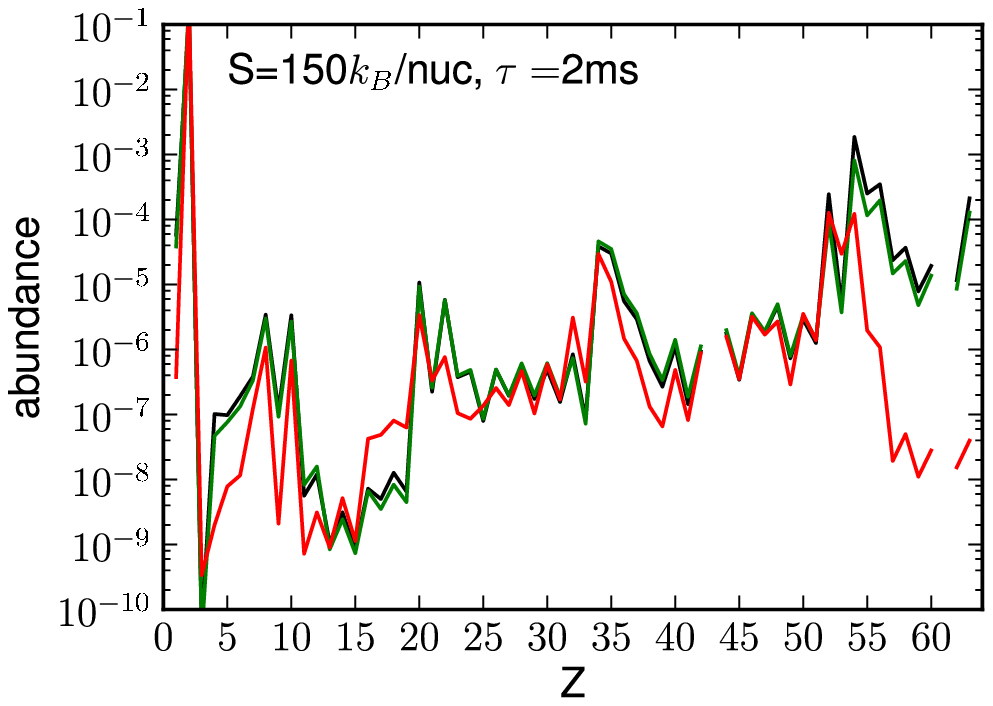}\\
  \includegraphics[width=6cm]{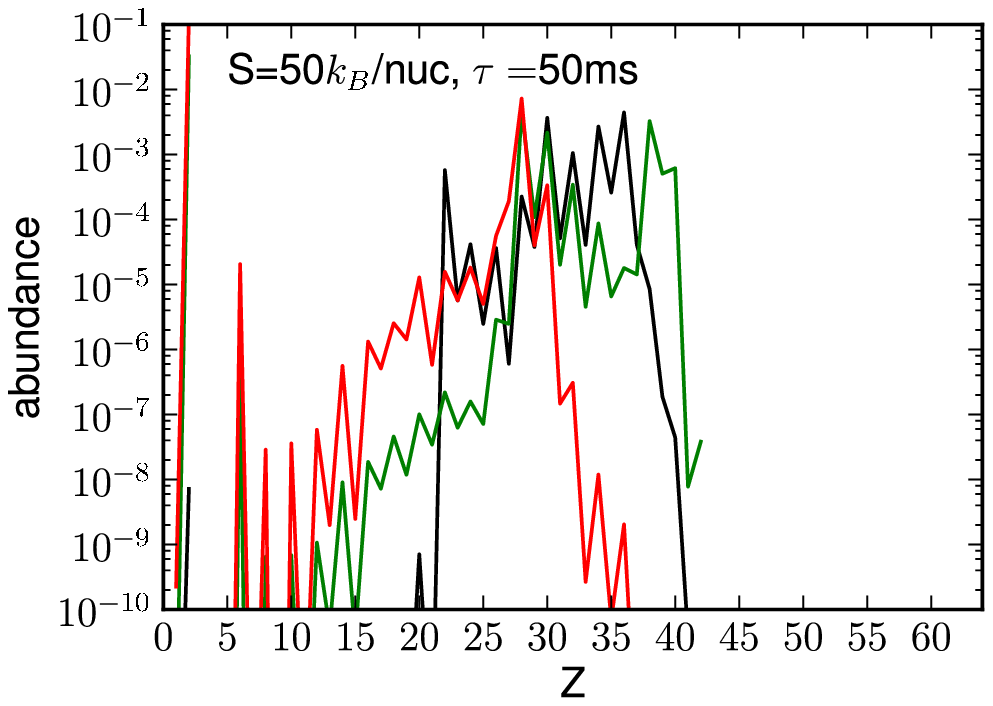}%
  \includegraphics[width=6cm]{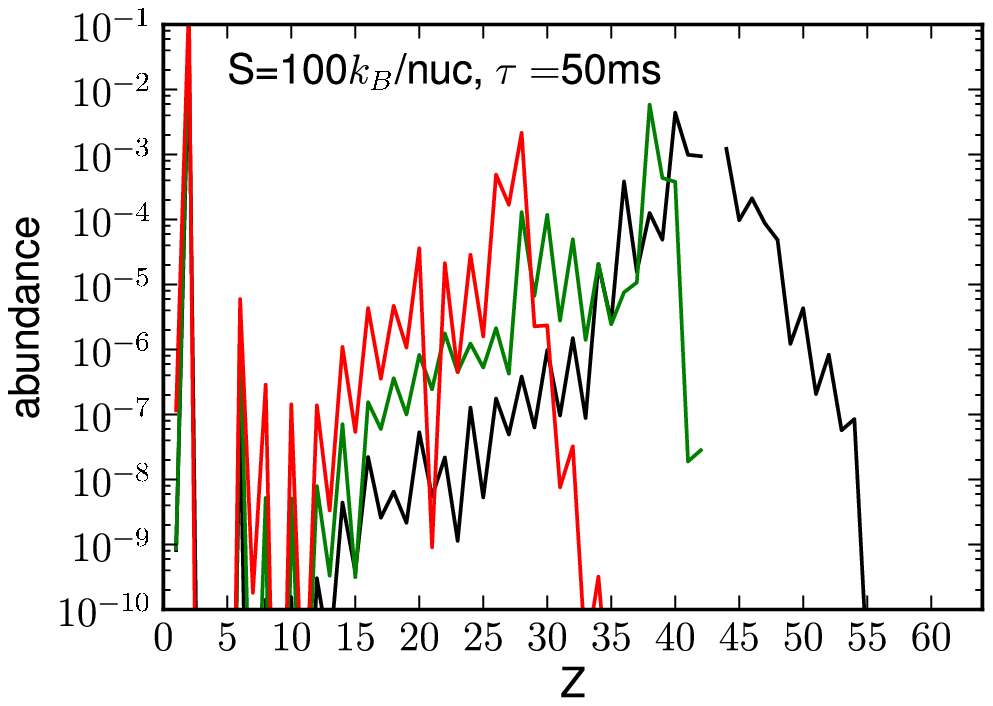}%
  \includegraphics[width=6cm]{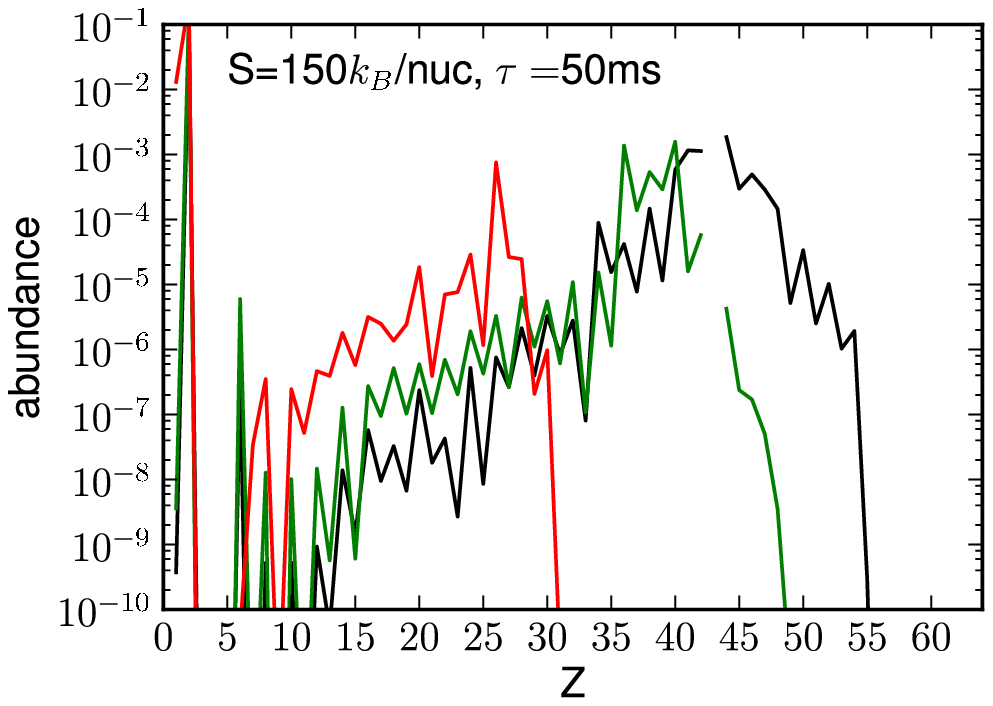}\\
  \caption{Elemental abundances for different electron fractions:
    $Y_e=$0.40 (black), 0.45 (green), 0.49 (red). The entropy and
    expansion time scale are given in the figures and are distributed
    as in Figure~\ref{fig:Xevol_nrich}.}
  \label{fig:YvsZ_nrich}
\end{figure}

Figure~\ref{fig:YvsZ_nrich} shows abundances corresponding to the
different panels of Figure~\ref{fig:Xevol_nrich}. This allows to link
the wind parameters, which determine the neutron-to-seed ratio, to the
elemental abundances and thus to the nucleosynthesis processes. For a
high entropy ($S>150 \, k_{\mathrm{B}}/\mathrm{nuc}$) and fast expansion
( $\tau<2$~ms) the r-process can form heavy elements. The other less
extreme cases produce elements up to different proton numbers. For
electron fraction close to 0.5 and a slow expansion with low entropy,
the abundances result from charged particle reactions. For
intermediate cases one talks about a weak r-process which reaches
$N=50$ (see Sect.~\ref{sec:weak_rprocess}). The last possibility,
which is not presented here but will be discussed later, is the $\nu
p$-process which builds heavy elements on the proton-rich side under
high neutrino fluxes (Sect.~\ref{sec:nup-process}).

\subsection{Additional ingredients}
\label{sec:add_ingredients}
The evolution of neutrino-driven winds is more complicated than the
simplified description of the previous section. The wind parameters
can become time-dependent due to an outer boundary, extra energy
sources, rotation and magnetic fields, or neutrino oscillations. There
have been several attempts of finding missing physical ingredients
which could lead to either an entropy increase, a shorter expansion
time scale, or reduction of the electron fraction. We discuss in this
section, the main additional ingredients that have been explored with
the goal of obtaining the r-process in neutrino-driven winds.

\vspace{0.5cm}

\noindent\emph{Wind termination}

\noindent The neutrino-driven wind expands through the early, slow
expanding supernova ejecta. When the wind becomes supersonic and the
early ejecta move slowly, the collision of both results in a wind
termination shock or reverse shock. This hydrodynamical feature has
been found in several supernova simulations \cite{Janka95, Janka96,
  Burrows.Hayes.Fryxell:1995, arcones.janka.scheck:2007,
  Arcones.Janka:2011, Fischer.etal:2010}.  Qian \& Woosley
(1996)~\cite{Qian.Woosley:1996} first used in a steady state model an
outer boundary with constant pressure at a radius corresponding to a
temperature of $\approx 2$~GK. The outcome was a slight increase of
the entropy and a reduction of the expansion time scale.  In
supersonic wind models \cite{Thompson.Burrows.Meyer:2001} an outer
boundary is necessary to decelerate the wind. In contrast in subsonic
winds or ``breeze'' \cite{Otsuki.Tagoshi.ea:2000,
  Wanajo.Kajino.ea:2001}, the wind velocity naturally decreases (see
Sect.~\ref{sec:wind_dyn}).  Other nucleosynthesis studies of the wind
termination also imposed an outer boundary pressure
\cite{Terasawa.Sumiyoshi.ea:2002} or kept the temperature constant
\cite{Wanajo.Itoh.ea:2002}.

The neutrino-driven wind was studied in detail for the first time by
means of modern, high-resolution, long-time hydrodynamical simulations
in Ref.~\cite{arcones.janka.scheck:2007}. Unlike the steady-state
models, where the wind expands freely, in reality the wind moves
through the early supernova ejecta. The interaction of the wind with
the slow ejecta is not a steady-state phenomenon, therefore
hydrodynamical simulations are required to study it in a consistent
way. The evolution of the wind termination shock or reverse shock was
investigated in Ref.~\cite{arcones.janka.scheck:2007} for different
stellar progenitors and various evolutions of the inner boundary
(i.e., contraction and compactness of the neutron star and neutrino
luminosity evolution). It was found that the reverse shock radius
($R_{\mathrm{rs}}$) depends on the wind mass outflow ($\dot{M}$) and
velocity ($v_w$), but also on the pressure of the slow, early
supernova ejecta:
\begin{equation}
  R_{\mathrm{rs}} \propto \sqrt{\frac{\dot{M}v_w}{P_{rs}}} \, .
  \label{eq:Rrs}
\end{equation}
The pressure in the supernova ejecta ($P_{rs}$) is linked to the
explosion energy ($E_{\mathrm{exp}}$) and to the progenitor, through
the shock radius ($R_s$) with: $P_{rs} \propto
E_{\mathrm{exp}}/R_{s}^3$. Moreover, convection and anisotropies have
a strong impact on this pressure and thus on the reverse shock radius
\cite{Arcones.Janka:2011}. When the supersonic neutrino wind collides
with the early supernova ejecta, the kinetic energy is transformed
into internal energy. The density and temperature of the shocked
matter are thus higher than in the wind (see
Fig.~\ref{fig:rs_prof}). The entropy at the reverse shock also
increases and can be estimated as~\cite{arcones.janka.scheck:2007}:
\begin{equation}
  s_{rs} \approx \left[ 
    \frac{s_w^{4/3}}{\beta^{1/3}} + 28.7
    \frac{R_{rs,8}^{2/3}v_{s,9}^{7/2}}{\dot{M}_{-5}^{1/3}}
  \right]^{3/4}\, .
  \label{eq:s_rs}
\end{equation}
Here $\beta$ is the relative jump in density at the reverse shock,
$R_{rs,8}$ the reverse shock radius in units of $10^8$~cm, $v_{w,9}$
the wind velocity in units of $10^9 \mathrm{cm}\,\mathrm{s}^{-1}$, and
$\dot{M}_{-5}$ the mass outflow normalized to $10^{-5}M_\odot$. If the
wind entropy ($s_w$) is much smaller than the entropy at the reverse
shock ($s_{rs}$), only the second term is relevant.  For a
10~$M_\odot$ progenitor, the entropy of the shocked matter can rise
above $400k_{\mathrm{B}} /
\mathrm{nuc}$~\cite{arcones.janka.scheck:2007}. For this low mass
progenitor the ram pressure is very small and the supernova shock
front expands very fast reaching large radii. This leads to a
reduction of the pressure in the early supernova ejecta ($P_{rs}$) and
to an increase of the reverse shock radius (Eq.~\ref{eq:Rrs}). This
and the larger wind velocity for this progenitor contribute to obtaining
high entropies at the position of the reverse shock
(Eq.~\ref{eq:s_rs}). However, this high entropy is attained only when the
temperature is already relatively low, $T_{rs}\approx 2-0.4$~GK. For a
successful r-process in neutrino-driven winds, the entropy has to be
high when the neutron-to-seed ratio is established. This occurs at
high temperature when charged-particle reactions are still building
seed nuclei at expenses of neutrons.

\begin{figure}[!h]
  \includegraphics[width=8cm]{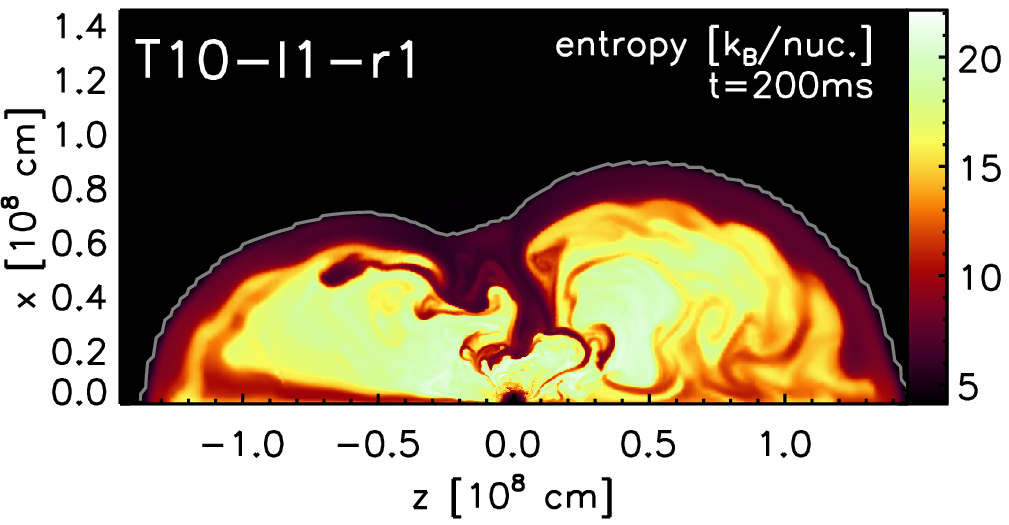}%
  \includegraphics[width=8cm]{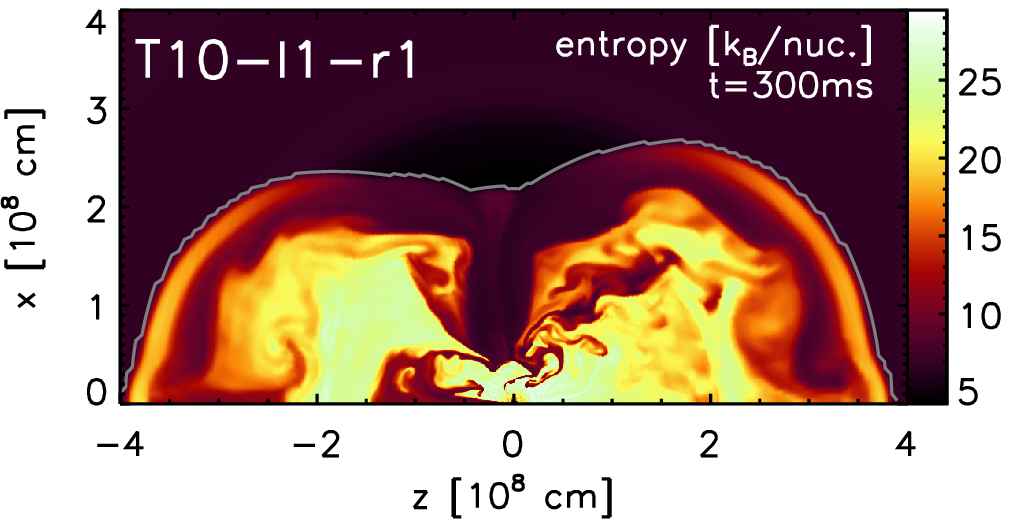}\\
  \includegraphics[width=8cm]{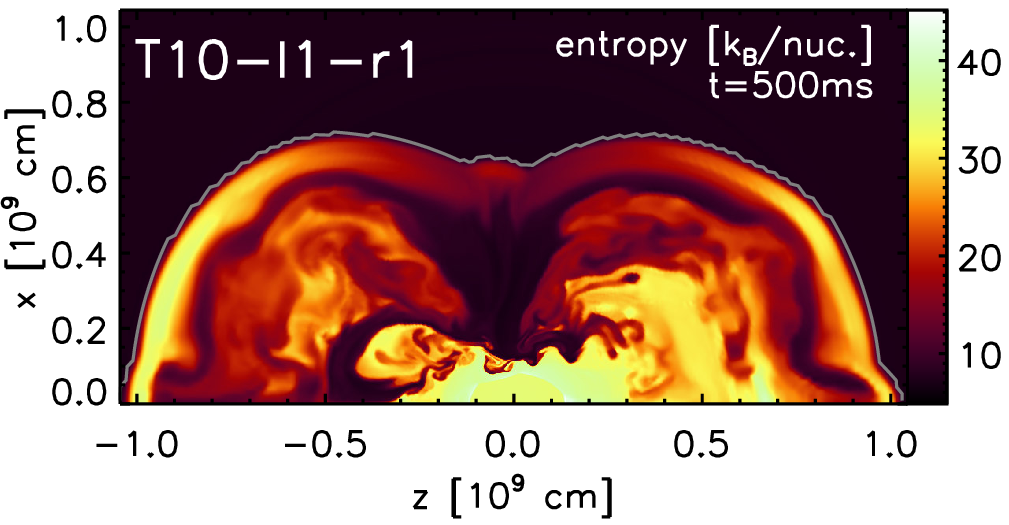}%
  \includegraphics[width=8cm]{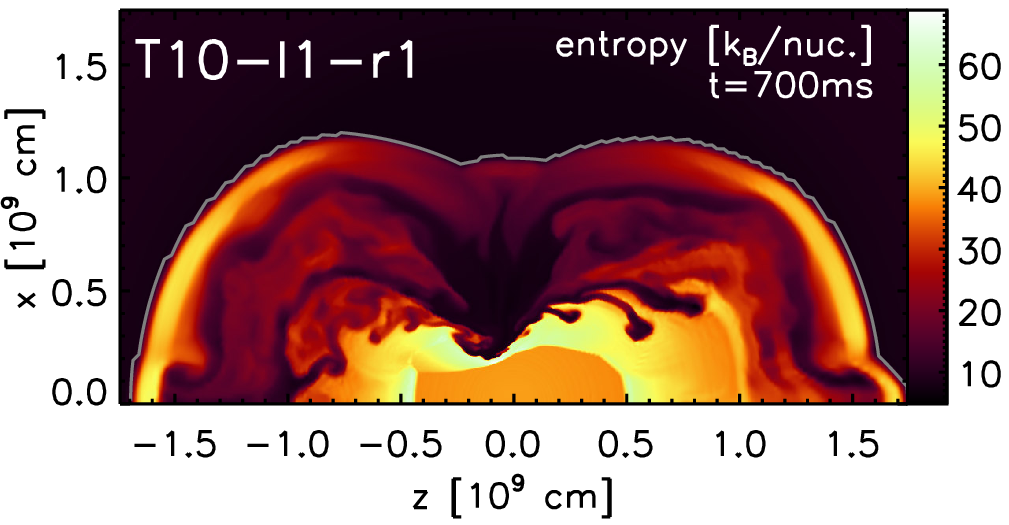}\\
  \includegraphics[width=8cm]{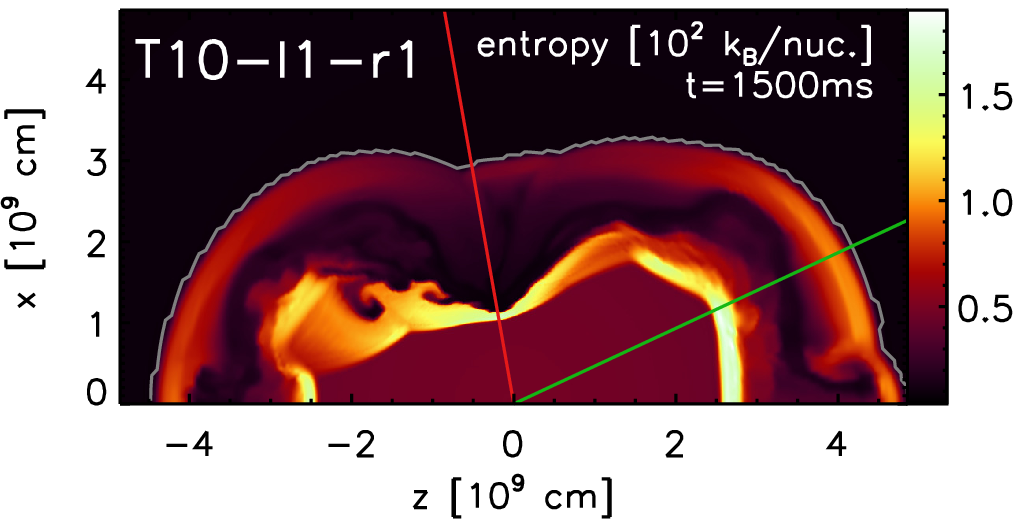}%
  \includegraphics[width=8cm]{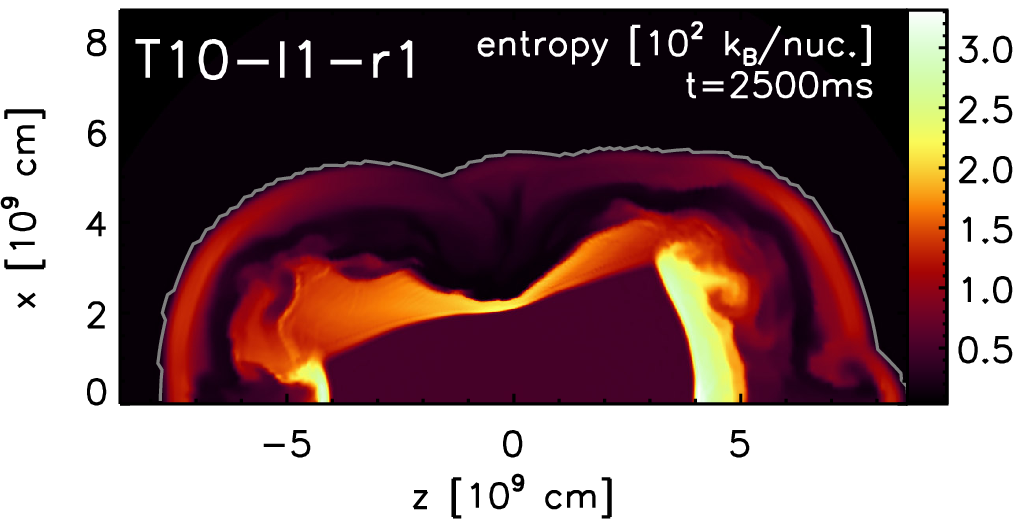}\\
  \caption{Entropy distribution in model
    T10-l1-r1~\cite{Arcones.Janka:2011} for different times after
    bounce as indicated in every panel.  The thin grey line marks the
    shock radius. In the panel for t = 1500 ms, the radial lines mark
    the angular directions at $\theta = 25$ degrees (green line) and
    100 degrees (red line), along which radial profiles are shown in
    Fig.~\ref{fig:rs_prof}.}
  \label{fig:2d_entr}
\end{figure}

\begin{figure}[!h]
  \includegraphics[width=7cm]{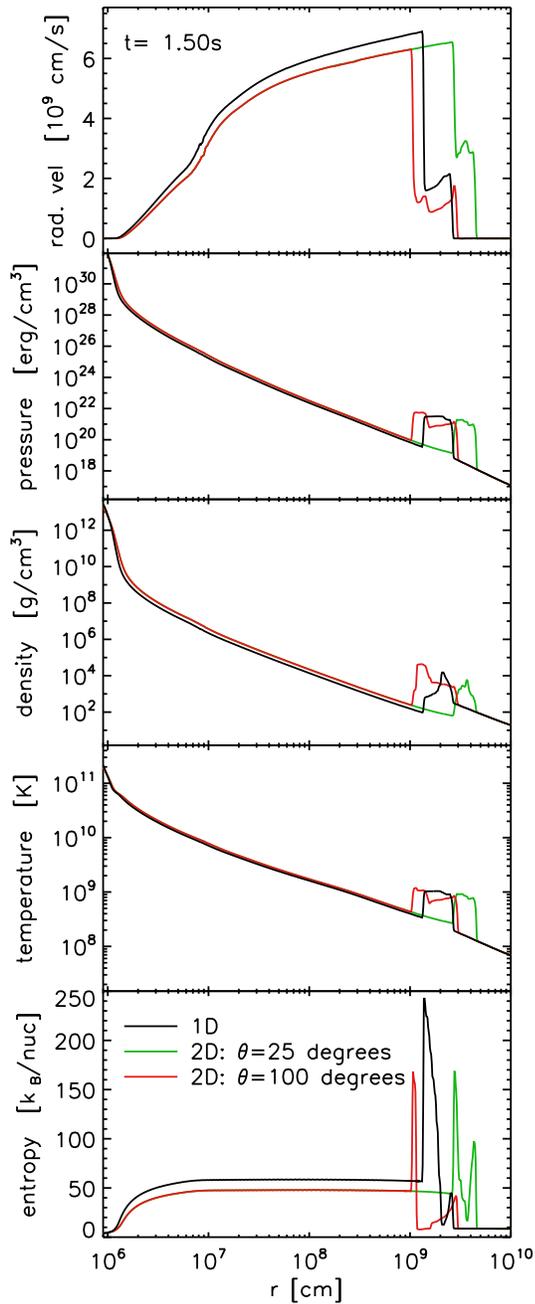}
  \caption{Comparison of one (black lines) and two-dimensional wind
    simulations showing the profiles of radial velocity, pressure,
    density, temperature, and entropy as functions of radius at time
    1.5 s after bounce. The one-dimensional model correspond to
    M10-l1-r1 described in Ref.~\cite{arcones.janka.scheck:2007}. For
    the two-dimensional model T10-l1-r1~\cite{Arcones.Janka:2011}, the
    profiles are shown at angles $\theta = 25$ degrees (green lines)
    and $\theta = 100$ degrees (red lines), corresponding to the lines
    of the same colors in Fig.~\ref{fig:2d_entr} (panel for
    $t=1500$~ms).}
  \label{fig:rs_prof}
\end{figure}

Multidimensional simulations \cite{Arcones.Janka:2011} indicate that
the wind stays spherically symmetric also in two-dimensional
simulations. This is expected as the neutrino emission from the
spherical neutrinosphere is isotropic in absence of rotation. However,
the anisotropic pressure distribution of the early ejecta has a big
impact on the position of the reverse shock and thus on the long-time
evolution. Figure~\ref{fig:rs_prof} shows the entropy distribution at
different times after bounce. The central region with constant entropy
(appearing at 700~ms) is the neutrino driven wind. Note the very
anisotropic form of the wind termination. The expansion after the wind
phase can be very different, therefore, if one uses parametric wind
models, it is recommended to vary the extrapolation (see
\ref{sec:par_rs}).

Different groups \cite{Kuroda.Wanajo.Nomoto:2008, Panov.Janka:2009,
  Arcones.MartinezPinedo:2011} have studied the impact of the reverse
shock assuming the r-process does occur in neutrino-driven winds. In
addition, the wind termination can also affect the $\nu p$-process
(\cite{Wanajo.Janka.Mueller:2011, Arcones.Frohlich:2011},
Sect.~\ref{sec:nup-process}).

\vspace{0.5cm}

\noindent\emph{Extra energy source}

\noindent An extra energy source can lead to an increase of the wind
entropy. Qian \& Woosley (1996)~\cite{Qian.Woosley:1996} explored this
possibility by artificially increasing the energy at different
positions. If the artificial energy source is close to the neutron
star, most of the energy goes into work against gravity without a
significant increase of entropy. If energy is deposited far out, in
regions where the temperature has dropped to 2~GK, then there may be
an impact on the r-process. If the extra energy source affects the
region where the mass outflow is determined ($\sim 15-35$~km), then
the entropy increases and the expansion time scale drops. Both
together are sufficient to have a successful r-process. The problem is
to identify such energy source.  Qian \& Woosley
(1996)~\cite{Qian.Woosley:1996} checked the uncertainties in neutrino
interactions and concluded that those are not sufficient to account
for the extra energy required. Wanajo (2006)~\cite{Wanajo:2006b}
showed that extra energy could be the outcome of an anisotropic
neutrino emission. Assuming a hot spot in the neutron star that emits
more neutrinos and using a simple model, r-process favourable
conditions were found. However, the origin of such a big anisotropy is
not yet clear. Further investigations based on multidimensional
simulations are necessary to understand this possibility better.

\vspace{0.5cm}

\noindent\emph{Rotation and magnetic fields}

\noindent Rotation and magnetic fields have been also included in
steady-state winds \cite{Nagataki.Kohri:2001, Thompson:2003,
  Suzuki.Nagataki:2005, Metzger.etal:2007}. Thompson
(2003)~\cite{Thompson:2003} studied the effect of strong dipole
magnetic fields and showed that matter is initially trapped by close
field lines. Later this matter escapes dynamically due to neutrino
heating and gets higher pressure and entropy, although the expansion
is slightly slower. For neutron-rich conditions ($Y_e<0.5$) and
magnetic fields of $B\gtrsim 10^{14}$~G the entropy increases
sufficiently to reach r-process favourable conditions. The amount of
matter ejected in these magnetic winds is higher ($\dot{M} \sim
10^{-4} M_\odot$) than in standard winds without magnetic
fields. However, these events are much too infrequent, in order to
fulfill galactic chemical evolution constraints. Suzuki \& Nagataki
(2005)~\cite{Suzuki.Nagataki:2005} showed that Alv\'en waves in strong
magnetized neutron stars \cite{Thompson.Duncan:1993} can accelerate
the wind and deposit energy. These waves can thus account for the
extra energy source suggested by Qian \& Woosley
(1996)~\cite{Qian.Woosley:1996} when the magnetic field reaches
$B\approx 3-5 \cdot 10^{14}$~G. Moreover, Metzger et
al. (2007)~\cite{Metzger.etal:2007} solved the magneto hydrodynamic
(MHD) wind equations for different rotation periods and magnetic
fields. With dipole fields of $B^{\mathrm{dip}} \gtrsim 10^{15}$~G and
rotation periods of $P\approx 10$~ms they were able to get closer to
the limiting conditions for a successful r-process
\cite{hoffman.woosley.qian:1997}. Although strong magnetic field and
fast rotation are not enough to produce r-process nuclei, MHD waves
can lead to sufficient energy to increase the entropy in agreement
with Ref.~\cite{Suzuki.Nagataki:2005}.

\vspace{0.5cm}

\noindent\emph{Neutrino oscillations}

\noindent All the additional ingredients discussed above bring the
neutrino-driven wind closer to a successful r-process but only if the
wind is neutron rich, i.e., $Y_e<0.5$. As explained in
Sect.~\ref{sec:weak_int}, there are still uncertainties on the
determination of $Y_e$. The r-process cannot occur in spherical
neutrino-driven winds unless it becomes sufficiently neutron rich. One
possibility to alter the electron fraction are neutrino
oscillations. These can increase the energy of electron antineutrinos
or/and reduce the electron neutrino fluxes. For the first to occur,
energetic neutrinos ($\nu_\tau$, $\nu_\mu$) need to oscillate into
electron neutrinos. This could occur via collective neutrino
oscillations \cite{Duan.Fuller.Qian:2010}.  However, the neutrino
spectra become very similar for all neutrino flavours during the wind
phase~\cite{Huedepohl.etal:2010} and thus no big increase of the
electron antineutrino energies is expected in this way. Notice that
this high energy could be important in proton-rich conditions for the
$\nu p$-process \cite{MartinezPinedo.etal:2011}.  Another exciting
possibility is due to matter-enhanced active-sterile oscillations as
it was suggested in Ref.~\cite{McLaughlin.etal:1999}. The existence of
sterile neutrinos remains an open question and it has been used to
explain anomalies in some detector
experiments~\cite{whitepaper2012}. This oscillation reduces the
electron neutrino flux at relative large radii, where the effective
neutrino heating has already occurred. McLaughlin et
al. (1999)~\cite{McLaughlin.etal:1999} showed that with the right
combination of neutrino parameters, electron neutrinos oscillate into
sterile neutrinos when $Y_e>0.3$, while when $Y_e<0.3$ antineutrinos
are the ones that oscillate. Similar discussions with sterile
neutrinos can be found in Refs.~\cite{Beun.etal:2006,
  Fetter.etal:2003, Hidaka.Fuller:2007}. Recently, a systematic study
has been performed \cite{Tamborra.etal:2012} based on a low mass
supernova progenitor. They found sterile neutrinos could mix with
active states and strongly impact $Y_e$. They considered also
collective active-active oscillations which become more significant at
later times and shift $Y_e$ towards higher values. Although there is
no clear proof of the existence of sterile neutrinos, there is still
room to explore the impact of this exotic phenomenon for the
neutrino-driven wind nucleosynthesis.

\section{Nucleosynthesis in neutrino-driven winds}
\label{sec:nuc_wind}
The nucleosynthesis processes that can occur in neutrino-driven winds
depend on the wind parameters. If the electron fraction is below 0.5,
there are two possibilities depending on the neutron-to-seed
ratio. High entropies and fast expansions lead to high
$Y_n/Y_{\mathrm{seed}} \gtrsim 100$ which allows the r-process to
build heavy elements up to uranium. When the conditions are less
neutron rich and the neutron-to-seed ratio is low
($Y_n/Y_{\mathrm{seed}}\lesssim 1$), elements up to $N=50$ are
synthesize in a weak r-process. Proton-rich winds are another
possibility which is favoured by current supernova models. In this
case the $\nu p$-process can produce elements heavier than $^{64}$Ge
along proton rich nuclei. The weak r-process and the $\nu p$-process
make neutrino-driven winds an exciting scenario to explain the origin
of lighter heavier elements such as Sr, Y, and Zr.

\subsection{r-process}
\label{sec:rprocess}
The extreme conditions required for the r-process are not found in
spherically symmetric neutrino-driven winds. However, high entropy
winds have been the center of many nucleosynthesis studies based on
steady state, parametric and hydrodynamic models. Here we summarize
the historical developments of the r-process in neutrino-driven winds.
Past investigations were also dedicated to analyze key aspect for the
r-process, e.g., nuclear physics input and impact of the dynamical
evolutions. These studies are critical to understand the r-process in
general for any astrophysical scenario.

\subsubsection{Historical overview:}
\label{sec:his}
In the 90's a new era started for the nucleosynthesis in core-collapse
supernovae and the developments of neutrino-driven wind studies. Since
1957 core-collapse supernovae have attracted interest as the
astrophysical site where heavy elements are synthesized by the
r-process \cite{Burbidge.Burbidge.ea:1957, Cameron:1957}. First
nucleosynthesis studies based on simple hydrodynamical simulations
were perform already by Hillebrandt et
al. (1976)~\cite{Hillebrandt.etal:1976}. These relied on prompt
supernova explosions where neutron rich matter from the outer layers
of the proto-neutron star is hydrodynamical ejected. Although the
solar system abundances could be approximately reproduced, the amount
of r-process material ejected in such events leads to an
overproduction of heavy elements if all core-collapse supernovae had
exploded in this way~\cite{Mathews.Cowan:1990}. The total mass
fraction of r-process material in our galaxy accounts for $\sim 10^4
M_\odot$ of the total $\sim 10^{11} M_\odot$. As the supernova rate is
$\sim 10^{-2} yr^{-1}$ \cite{Qian.Wasserburg:2008} and the age of the
galaxy when the solar system formed was $10^{10}yr$, the r-process
material produced per supernova in the history of the galaxy can be
only $\lesssim 10^{-4} M_\odot$, assuming the r-process takes place in
every supernovae. Therefore, Hillebrandt et al. (1976)
\cite{Hillebrandt.etal:1976} argued that the prompt explosions rich in
r-process material are rare events. By now we know that the prompt
shock bounce after the collapse is not sufficient to lead to a
successful supernova explosion but occurs by some delayed
mechanism~\cite{Myra.Bludman:1989, Bethe:1990, Kitaura06,
  Fischer.etal:2010}.

First delayed neutrino-driven explosions were reported by Bethe \&
Wilson (1985) \cite{Bethe85} and followed by many nucleosynthesis
studies. Two groups studied the late hydrodynamical evolution based on
Wilson explosions.

Woosley \& Hoffman (1992) \cite{Woosley.Hoffman:1992} studied the
nucleosynthesis from NSE to the alpha-rich freeze-out of charged
particle reactions in the high entropy wind from the nascent
proto-neutron star.  They suggested that the later evolution could
lead to high entropy and low electron fractions and thus to a
successful r-process.  In a following paper, Meyer et
al. (1992)~\cite{Meyer.Mathews.ea:1992} followed the evolution and
nucleosynthesis after the alpha-rich freeze-out based on different
values of $Y_e$. The final abundances, based on a superposition of
various trajectories with different neutron excess, agreed rather well
with the solar system abundances. Finally, Woosley et al. (1994)
\cite{Woosley.etal:1994} performed hydrodynamical simulations with the
spherically symmetric radiation hydrodynamic code of Wilson \& Mayle
(1993)~\cite{Wilson.Mayle:1993} for a 20$M_\odot$ progenitor and
followed the evolution of the ejecta for 20~s after the
explosion. Already 5~s after bounce appropriate conditions for the
r-process were found in the $10^{-4}M_\odot$ ejected by the wind. A
nice agreement with solar system r-process abundances was obtained
only with a small overproduction around $A=90$.  The success of this
work was summarized at the end of their paper: ``... the problem is in
need of further study, but we are gratified to have found what seems
to be the most promising site yet proposed for the production of the
r-process elements''.

However, already in the same year these results could not be
reproduced by an independent group. Witti et al. (1994)
\cite{Witti.etal:1994} used also a Wilson supernova model for a
25$M_\odot$ star and studied the late evolution starting at 0.6~s
after bounce. They found entropy values below $100 \, k_{B}
\mathrm{nuc}^{-1}$ and electron fractions above 0.45, compared to
$S\sim 400 \, k_{B} \mathrm{nuc}^{-1}$ and $Y_e\sim 0.35$ reported by
Woosley et al. (1994)~\cite{Woosley.etal:1994}. In their first paper
\cite{Witti.etal:1994}, the moderate entropy of their models resulted
in an alpha-rich freeze-out with too many seed nuclei and too few
neutrons. This led, similar to Ref.~\cite{Woosley.Hoffman:1992}, to a
strong overproduction of nuclei around $A=90$, related to the neutron
magic number $N=50$. Their second paper \cite{Takahashi.etal:1994}
presents the possibility of having an r-process in the neutrino-driven
winds if the entropy is assumed to be larger. Increasing the entropy
by a factor 5.5 (reducing the density by the same factor, $S\propto
T^3/\rho$) reproduced the solar system abundances and solved the
overproduction problem for $A\sim 90$. They concluded that ``the
neutrino wind in core-collapse supernovae is a very promising site for
the r-process nucleosynthesis'', but still ``much remains to be worked
out''.

These both works presented the exciting possibilities for the
r-process in neutrino-driven winds and identified the three key
parameters: entropy, expansion time scale, and electron
fraction. However, the results did not agree and were far from being
final. Further detailed studies of the wind and supernovae were
necessary and have emerged since then.

Qian \& Woosley (1996)~\cite{Qian.Woosley:1996} studied the wind based
on an analytic model and could not find appropriate conditions for an
r-process in general, therefore investigated the impact of missing
ingredients, such as an outer boundary or an extra energy source (see
Sect.~\ref{sec:add_ingredients}). They concluded that any effect that
strengthens the gravitational potential leads to an increase of the
entropy.  Following this line, Cardall \& Fuller
(1997)~\cite{Cardall.Fuller:1997} included a general relativity
treatment of the wind. The outcome of this study was a more compact
neutron star and, consequently, higher wind entropy than in the
Newtonian case. Hofmann et al. (1997)~\cite{hoffman.woosley.qian:1997}
generalized the model of Qian \& Woosley (1996) \cite{Qian.Woosley:1996}
and constrained the wind parameters.  These investigations showed
again that for the electron fractions and dynamical time scales
reported by Woosley et al. (1994)~\cite{Woosley.etal:1994}, the entropy
cannot reach values of $\sim 400 \, k_{\mathrm{B}}/\mathrm{nuc}$.

More advanced studies were performed by Otsuki et
al. (2000)~\cite{Otsuki.Tagoshi.ea:2000} and Thompson et
al. (2001)~\cite{Thompson.Burrows.Meyer:2001}.  These two detailed
studies are independent of the supernova mechanism and thus provide
general properties. They clearly demonstrated that general relativity
increases the wind entropy (up to $ \sim 40\%$) compared to Newtonian
approach. Including such corrections in the neutrino treatment has a
minor effect on the wind \cite{Otsuki.Tagoshi.ea:2000}: the bending of
the neutrino trajectories increases the energy deposition ($\dot{q}$),
while the redshift reduces it.

\begin{figure}[!htb]
  \includegraphics[width=8cm,angle=90]{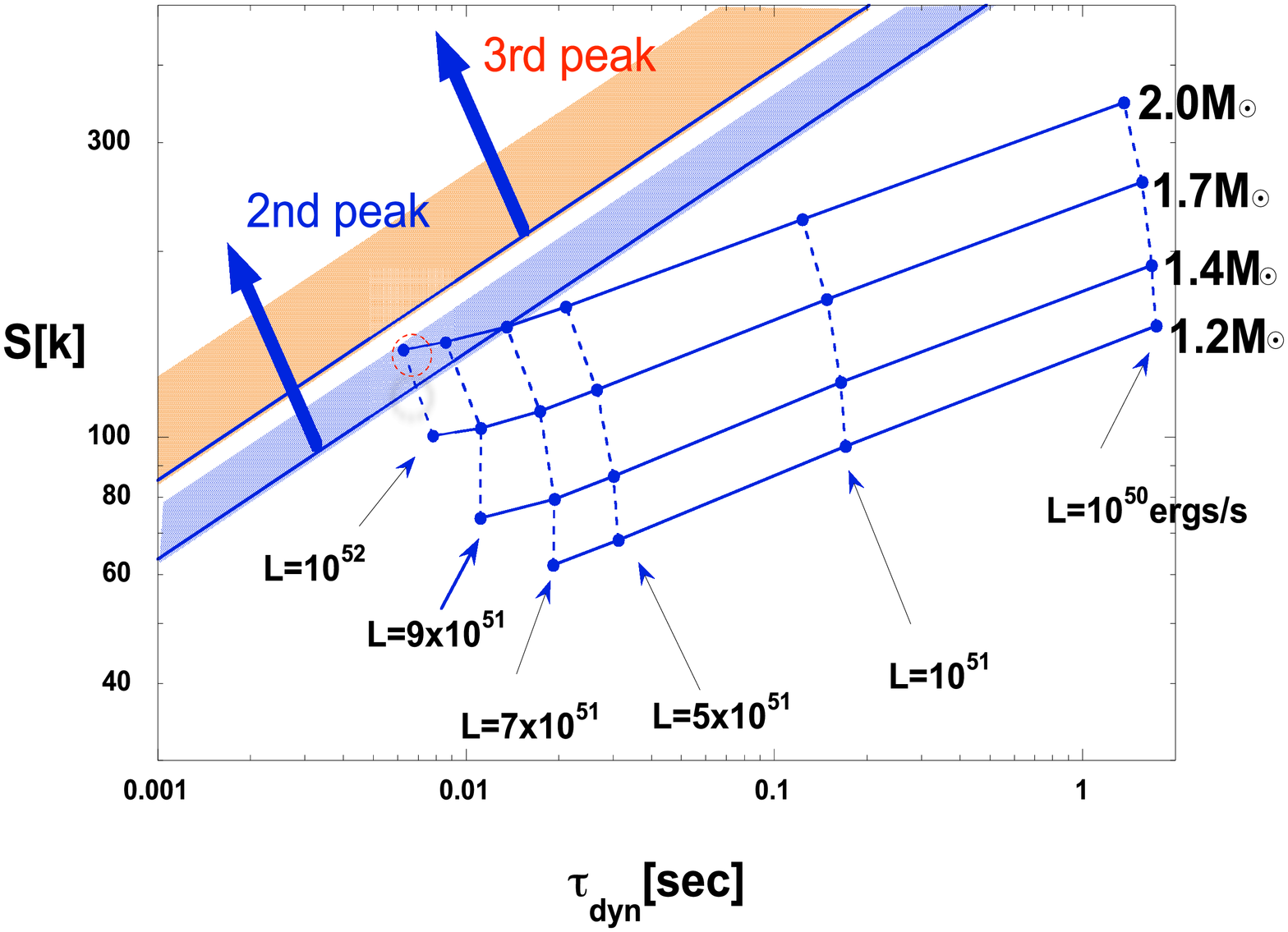}
  \caption{Figure courtesy of Kaori Otsuki. Every point corresponds to
    the solution of the steady wind
    equations~\cite{Otsuki.Tagoshi.ea:2000} for given neutrino
    luminosities and neutron star masses as indicated in the
    figure. The electron fraction is $Y_e=0.45$. The shadowed regions
    represent the necessary entropies and expansion time scales to
    obtain the second and third r-process peaks.}
  \label{fig:kaori}
\end{figure}

We use here Fig.~\ref{fig:kaori} to illustrate which combination of
wind parameters favour the r-process based on the steady-state models
of Otsuki et al. (2000)~\cite{Otsuki.Tagoshi.ea:2000}. Different
neutron star masses (connected by solid lines) and various
luminosities (connected by dashed lines) lead to different solutions
in the entropy--time scale plane for $Y_e=0.4$. The color shadowed
regions indicated the necessary conditions for the synthesis of
elements up to the second and third r-process peak. Notice that this
is possible only with very fast expansions (small dynamical time
scale, $\tau_{\mathrm{dyn}} <0.03$~s) and high entropies ($S=150-300
\, k_{\mathrm{B}}/\mathrm{nuc}$). This is achieved
\cite{Otsuki.Tagoshi.ea:2000, Thompson.Burrows.Meyer:2001} if the
neutron star is very compact (i.e., a massive neutron star with small
radius).  Figure~\ref{fig:kaori} clarifies the impact of the
luminosity (Eqs.~\ref{eq:swind_qw} and~\ref{eq:tau_qw}): lower
luminosities lead to higher entropies but also slower expansion.

All these efforts, investigating the neutrino-driven winds with
parametric models, did not lead to conditions which reproduce the
r-process as reported in Woosley et al. (1994)
\cite{Woosley.etal:1994}. Moreover, the first long-time hydrodynamic
simulations \cite{arcones.janka.scheck:2007}, which could follow the
wind during seconds with enough resolution, confirmed the results of
steady state models even in two-dimensional
models~\cite{Arcones.Janka:2011}. These simulations were performed
with simplified neutrino transport, therefore the electron fraction
obtained is not very exact.  More sophisticated simulations using
Boltzmann neutrino transport ~\cite{Huedepohl.etal:2010,
  Fischer.etal:2010} indicate that the neutrino-driven wind may be
proton-rich in contrast to the very neutron-rich condition found by
Woosley et al. (1994)~\cite{Woosley.etal:1994}. Therefore, spherically
symmetric neutrino-driven winds are not a reliable candidate to
explain the origin of heavy elements via the r-process. However, the
wind ejecta being slightly neutron rich or proton rich remain an
exciting scenario for other nucleosynthesis processes
(Sect.~\ref{sec:lepp}-\ref{sec:nup-process}).

\subsubsection{Impact of wind dynamical evolution on the r-process:}
\label{sec:dynevol}

If one assumes that the r-process occurs in high entropy
neutrino-driven winds there are two important aspects that strongly
affect the abundance of heavy elements: long-time dynamical evolution
and nuclear physics input.

Most of the r-process studies in neutrino-driven winds have focused in
finding the appropriated conditions to obtain a high neutron-to-seed
ratio, which is determined in the early phase of the evolution during
NSE and charged-particle reactions. However, the final r-process
abundances (that one wants to compare to observation) are also
affected by the long-time evolution when the temperature has already
dropped to $T=2-0.1$~GK. The wind termination has a strong impact on
the evolution of the temperature at late times
(Sect.~\ref{sec:add_ingredients}).  It was Wanajo
(2007)~\cite{Wanajo:2007} who defined two types of r-process depending
on the dynamical evolution: a hot and a cold r-process. In the hot
r-process the reverse shock is at relative high temperatures
($T\approx1$~GK), there is an $(n,\gamma)-(\gamma,n)$ equilibrium and
beta-decay time scales are longer than the neutron capture and
photo-dissociation time scales. For the cold r-process, the
photodissociation time scale becomes longer once the temperature drops
below ∼ 0.5 G and the evolution proceeds by a competition between beta
decay and neutron capture \cite{Blake.Schramm:1976, Panov.Janka:2009}.

The dependence of the abundances on the long-time
evolution~\cite{Arcones.MartinezPinedo:2011, Farouqi.etal:2010,
  Kuroda.Wanajo.Nomoto:2008} is due to the different nuclear reactions
that become important for every evolution. Therefore, the impact of
the nuclear physics will also depend on the evolution as shown in
detailed in Refs.~\cite{Arcones.MartinezPinedo:2011,
  Farouqi.etal:2010}.

\subsubsection{Impact of nuclear physics input on the r-process:}
\label{sec:nucphysinput}

In addition to the difficulties to find the astrophysical scenario
where half of the heavy elements are produced in the
universe~\cite{Thielemann.etal:2011, arnould.goriely.takahashi:2007},
only few of the very exotic nuclei involved in the r-process have been
produced up to now in current rare isotope
facilities~\cite{Kratz.etal:2007b}. Therefore, nucleosynthesis
calculations rely on theoretical predictions towards extreme
neutron-rich nuclei far from stability.  There are several nuclear
physics inputs with a strong impact on the final abundances.

Nuclear masses determine the energy thresholds for all relevant
reactions: neutron capture, photodissociation, and beta decay. Their
impact on the abundances have been studied (see
e.g.,~\cite{Arcones.MartinezPinedo:2011, Farouqi.etal:2010,
  Kratz.Bitouzet.ea:1993, Arcones.Bertsch:2012}) by using different
theoretical mass models (e.g., FRDM \cite{Moeller.etal:1995}, ETFSI
\cite{Aboussir.etal:1995}, Duflo \& Zuker \cite{Duflo.Zuker:1995},
HFB-x \cite{Goriely.etal:2009, Goriely.etal:2010}). Beta decay
rates~\cite{Moeller.etal:2003} control the speed of the r-process and
during decay to stability become specially important due to the
delayed neutron emission.

Reaction rates are also critical to understand the r-process. Neutron
capture rates \cite{Rauscher.Thielemann:2004} have been shown to have
a non-negligible impact on the final abundances as they shift the
position of peaks and affect the rare-earth peak ($A \approx
160$)~\cite{Surman.Engel.ea:1997, surman.etal:2009,
  Arcones.MartinezPinedo:2011, Mumpower.etal:2012}.  Reactions
involving neutrinos have been discussed extensively within the
neutrino-driven wind models. Neutrino absorption on nuclei can mimic a
beta decay and accelerates the r-process~\cite{Fuller.Meyer:1995,
  Mclaughlin.etal:1996}, and also cause neutrino-induced neutron
emission~\cite{Qian.etal:1997, Haxton.etal:1997} which can produce
rare species from neighbouring nuclei with high abundances. While
neutrino absorption can speed up the r-process by mimicking $\beta^-$
decays, it can also weaken an r-process by disintegrating alpha
particles, which can then contribute to a larger seed nuclei
population and thus a decreased neutron-to-seed
ratio~\cite{Meyer:1995}.

Fission can play an important role specially when several fission
cycles occur~\cite{Beun.etal:2006, Beun.etal:2008, Panov.etal:2005,
  Panov.etal:2010}, leading to a robust r-process abundance
pattern. The main fission channels include neutron induced fission,
spontaneous fission, and beta-delayed fission~\cite{Mamdouh.etal:1998,
  Myers:1999, Goriely.Hilaire.etal:2009,
  Panov.etal:2005}. Neutrino-induced fission has been investigated in
Ref.~\cite{Kolbe.etal:2004}, but found to be negligible in
neutrino-driven wind environments~\cite{Thielemann.etal:2007,
  MartinezPinedo.etal:2007}.

A more detailed discussion of the impact of these nuclear physics
inputs on the r-process is beyond the scope of this review.

\subsection{Lighter heavy elements: Sr, Y, Zr}
\label{sec:lepp}

Although spherically symmetric neutrino-driven winds are not a good
candidate for r-process, they are an exciting possibility to explain
the origin of lighter heavy elements, such as Sr, Y, Zr. There are
observational indications that several components or sites contribute
to the so-called r-process elements. The r-process component results
from subtracting the s-process component to the solar system
abundances. The r-process is thus not consistently obtained by
integrating the contribution of different sites within a galactic
chemical evolution model. Therefore, the r-process component extracted
from the solar system is not a real process but includes residual
abundances left after removing the s-process. Moreover, observations
of UMP stars \cite{Sneden.etal:2008} and meteorites
\cite{Wasserburg.etal:1996} indicates that indeed several
nucleosynthesis processes have contributed to build the abundances of
the ``residual r-process''.

The elemental abundances observed in the atmosphere of UMP stars
\cite{Sneden.etal:2008, Johnson.Bolte:2002} present a robust pattern
for heavy elements $56<Z<83$ in agreement with the solar r-process
component. For few stars it has been possible to observe also
Tellurium showing a robust pattern even for the second r-process
peak~\cite{Roederer.etal:2012}. In contrast, the abundances of lighter
heavy elements $Z<47$ show some scatter which points to the existence
of at least two primary processes \cite{Sneden.etal:2008}.
Observations suggest that there could be even three components or
processes: 1) robust heavy r-process, 2) Sr, Y, Zr, 3) Ag and
Pd~\cite{Hansen.Primas:2011}.

The origin of lighter heavy elements (Sr, Y, Zr) was investigated by
Travaglio et al (2004)~\cite{Travaglio.Gallino.ea:2004} with their
s-process calculation combined with a galactic chemical evolution
model.  The abundances of several isotopes ($^{86}$Sr, $^{93}$Nb,
$^{96}$Mo, $^{100}$Ru, $^{104}$Pd, $^{110}$Cd) could not be explained
by adding s- and r-process (as observed in UMP stars)
contributions. They suggested that these isotopes are produced by the
Lighter Element Primary Process (LEPP), but they did not specify the
astrophysical site nor the nuclear reactions involved. Other authors
had proposed before that the process producing elements with $A<130$
was a weak r-process ~\cite{Truran.Cowan:2000} or charged-particle
reactions ~\cite{Woosley.Hoffman:1992, Freiburghaus.Rembges.ea:1999,
  Qian.Wasserburg:2007}. Montes et al. (2007)~\cite{Montes.etal:2007}
showed that there is an anticorrelation between heavy r-process and
Sr-like elements in UMP stars. The elemental abundances of the Sr-like
elements in UMP (stellar LEPP) agrees with the missing component in
the solar system~\cite{Travaglio.Gallino.ea:2004} (solar LEPP), within
the observational error bars (see abundances of HD~122563 in Fig.~5 of
Ref.~\cite{Montes.etal:2007}).  However, it remains an open question
whether both are due to the same nucleosynthesis process.

\begin{figure}[!htb]
  \includegraphics[width=8cm]{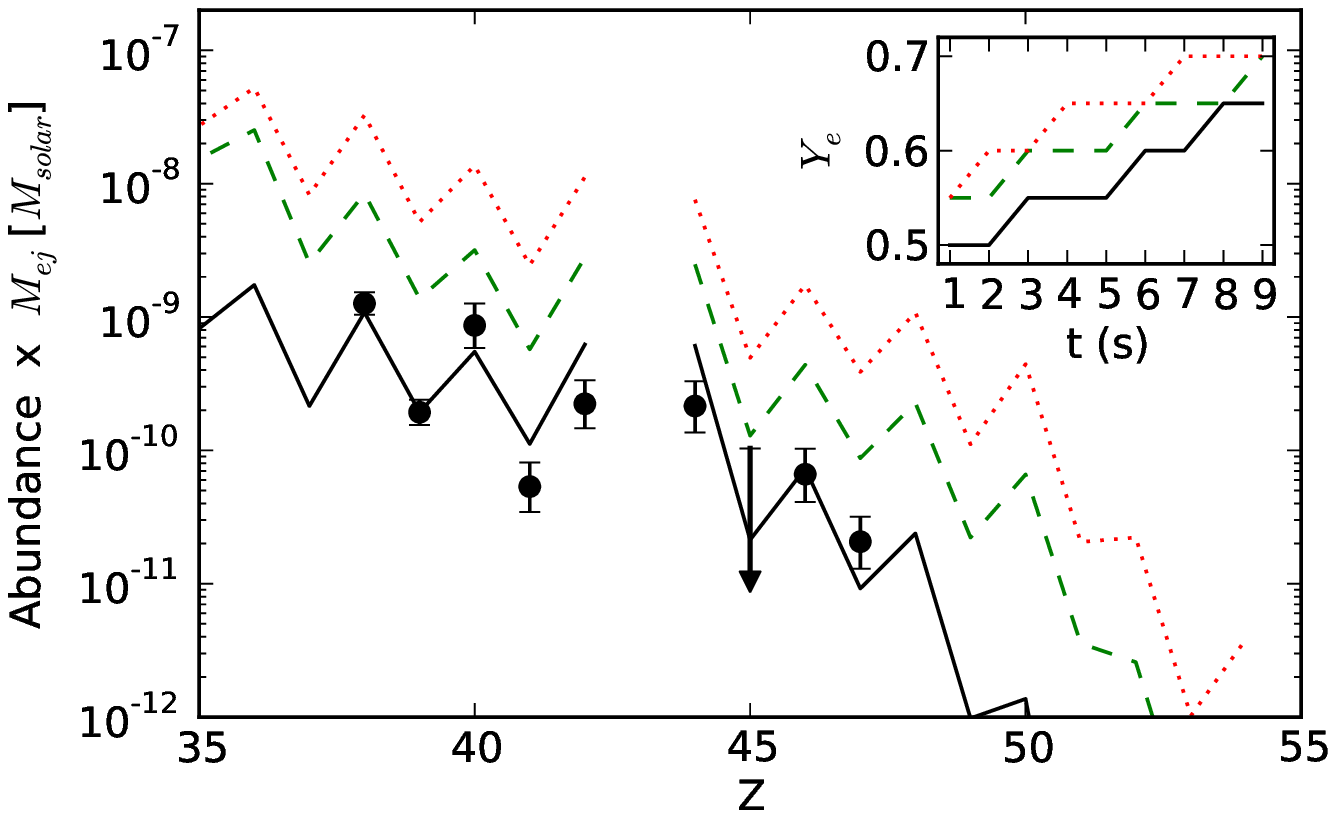}
  \includegraphics[width=8cm]{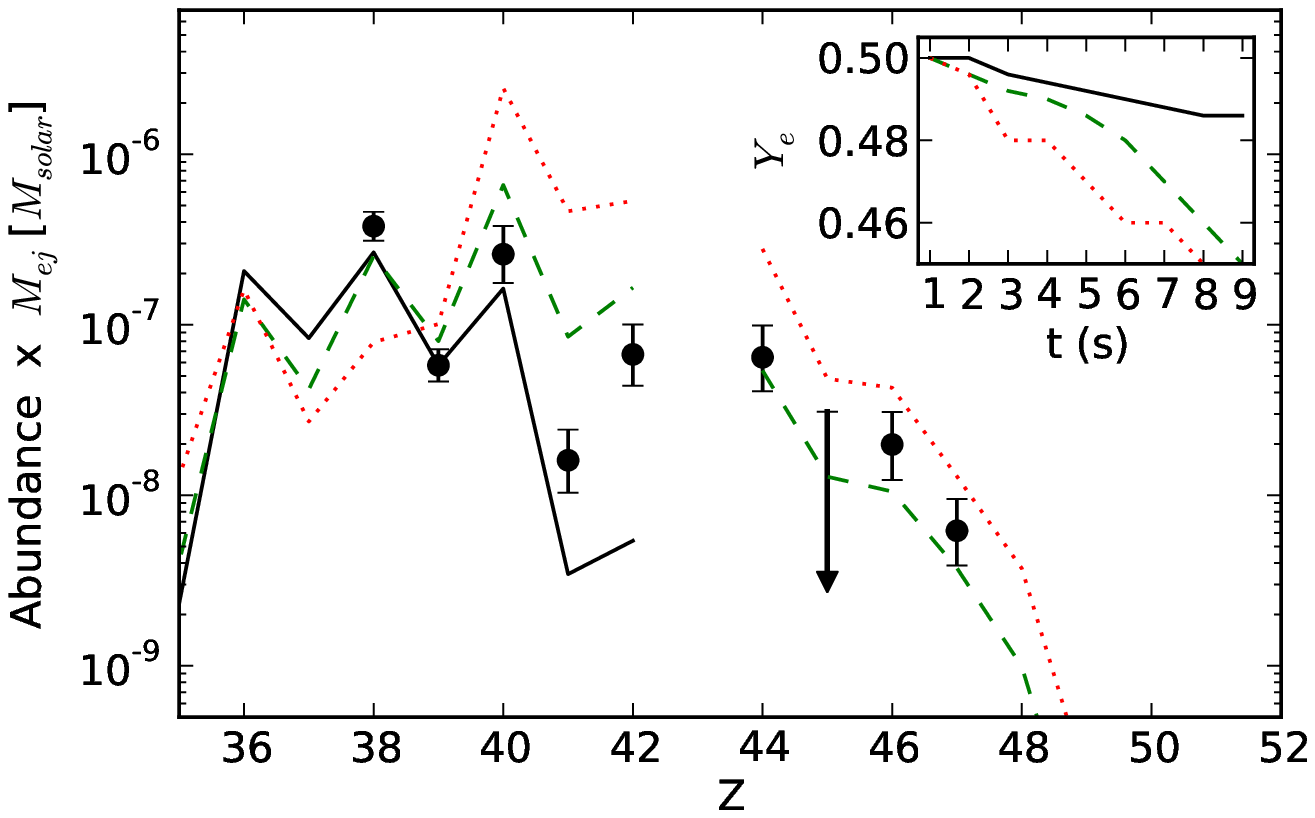}
  \caption{Elemental abundances for a superposition of mass zones with
    different electron fractions which are shown in the inset as a
    function of time after core collapse. The observed
    abundances~\cite{Honda.etal:2004, Honda.etal:2006,
      Montes.etal:2007} are shown by dots and rescaled to fit the
    solid line abundances.}
  \label{fig:lepp}
\end{figure}

Based on the observations of UMP stars, Qian and Wasserburg have
developed a phenomenological model \cite{Qian.Wasserburg:2001,
  Qian.Wasserburg:2007, Qian.Wasserburg:2008} to explain the
astrophysical site and nucleosynthesis process contributing to the
residual solar r-process abundances. In their model the lighter heavy
elements are produced by charged-particle reactions in neutrino-driven
winds. This has been confirmed~\cite{Arcones.Montes:2011} with
hydrodynamical simulations of core-collapse supernovae and
neutrino-driven winds using a parametrized electron fraction
evolution.  Roberts et al. (2010)~\cite{Roberts.etal:2010} calculated
also the integrated nucleosynthesis based on hydrodynamical
simulations of the neutrino-driven wind. Figure~\ref{fig:lepp} shows
the elemental abundances based on neutrino-driven wind simulations and
compared to observations. Note that different $Y_e$ evolutions are
employed to explore possible uncertainties in the neutrino physics.
For neutron- and proton-rich conditions, LEPP elements are produced
with different isotopic abundances. In neutron-rich winds, the
composition is dominated by neutron-rich isotopes and there is even an
overproduction for $A=90$ (related to the magic number $N=50$).  In
proton-rich winds the elemental-abundance pattern is quite robust
against small variations of the wind parameters and the isotopic
composition is characterized by neutron-deficient isotopes, i.e. those
on the left side of stability. In the following sections we discuss
the nucleosynthesis processes leading to these different isotopic
abundances.

\subsection{Weak r-process}
\label{sec:weak_rprocess}

Most recent simulations predict proton-rich conditions, however there
are still uncertainties on the neutrino physics which could lead to
slightly neutron-rich conditions (Sect.~\ref{sec:weak_int},
\ref{sec:add_ingredients}, \cite{MartinezPinedo.etal:2012,
  Roberts:2012, Roberts.Reddy:2012}). Moreover, two-dimensional
simulations of the explosion of an ONeMg progenitor indicates that
small neutron-rich pockets can be ejected because the expansion is
very fast and neutrinos have not sufficient time to change their
neutron richness~\cite{Wanajo.Janka.Mueller:2011}. Similar conditions
are found in explosion based on an unrealistic quark phase
transition~\cite{Nishimura.etal:2012}.

The dependence of the weak r-process abundances on wind parameters
have been investigated based on classical r-process
models~\cite{Kratz.etal:2007}, which assume
$(n,\gamma)-(\gamma,n)$-equilibrium, and on dynamical parametric high
entropy wind models~\cite{Farouqi.etal:2009}. These kind of studies
aim to identify the astrophysical conditions which lead to good
agreement between calculated and observed abundances.  We have shown
in Sect.~\ref{sec:nuc_par} the composition of slightly neutron-rich
wind based on the parametric wind trajectories presented in
\ref{sec:par_traj}.  The abundance dependence on wind parameters
(Fig.~\ref{fig:YvsZ_nrich}) can be understood by looking at the
evolution of neutrons and seed nuclei in
Fig.~\ref{fig:Xevol_nrich}. There are several combinations of wind
parameters ($s$, $Y_e$, $\tau$) which produce the lighter heavy
elements via a weak r-process and small variations of these parameters
leads to different final abundances as shown in
Ref.~\cite{Arcones.Montes:2011} for the LEPP.

\begin{figure}[!htb]
  \includegraphics[width=8cm]{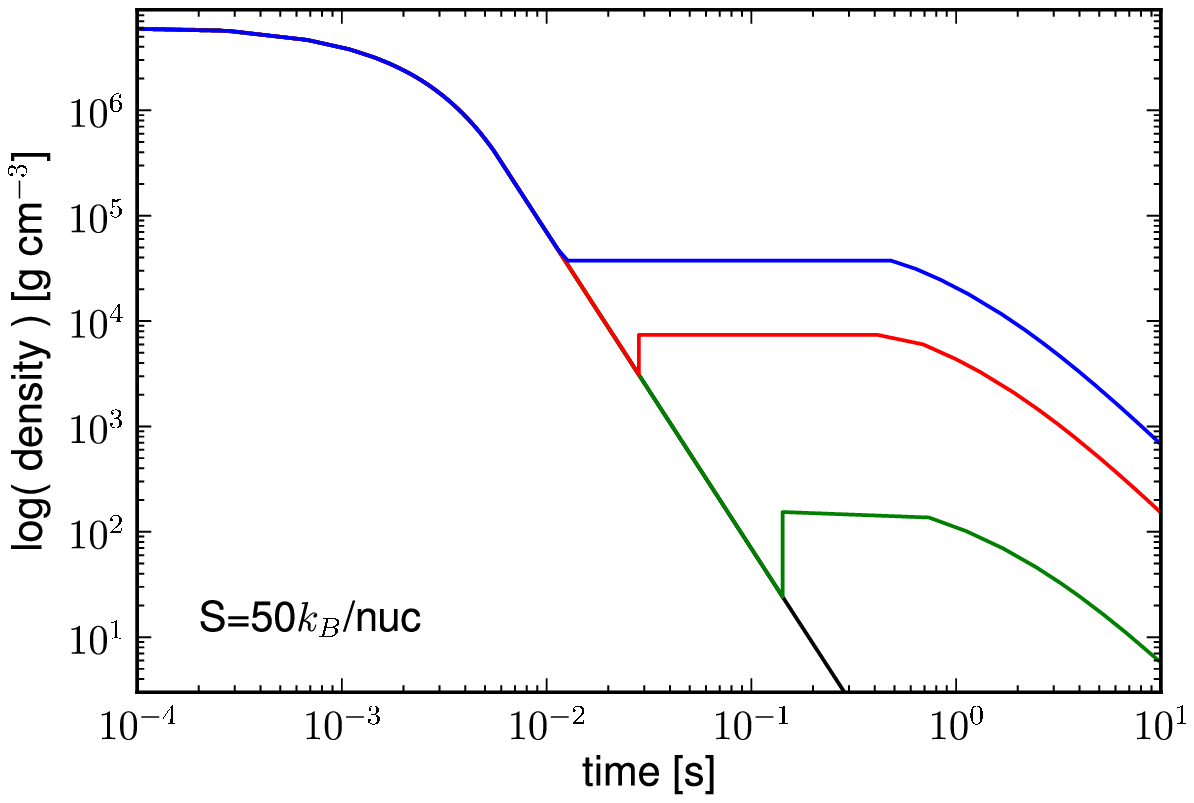}%
  \includegraphics[width=8cm]{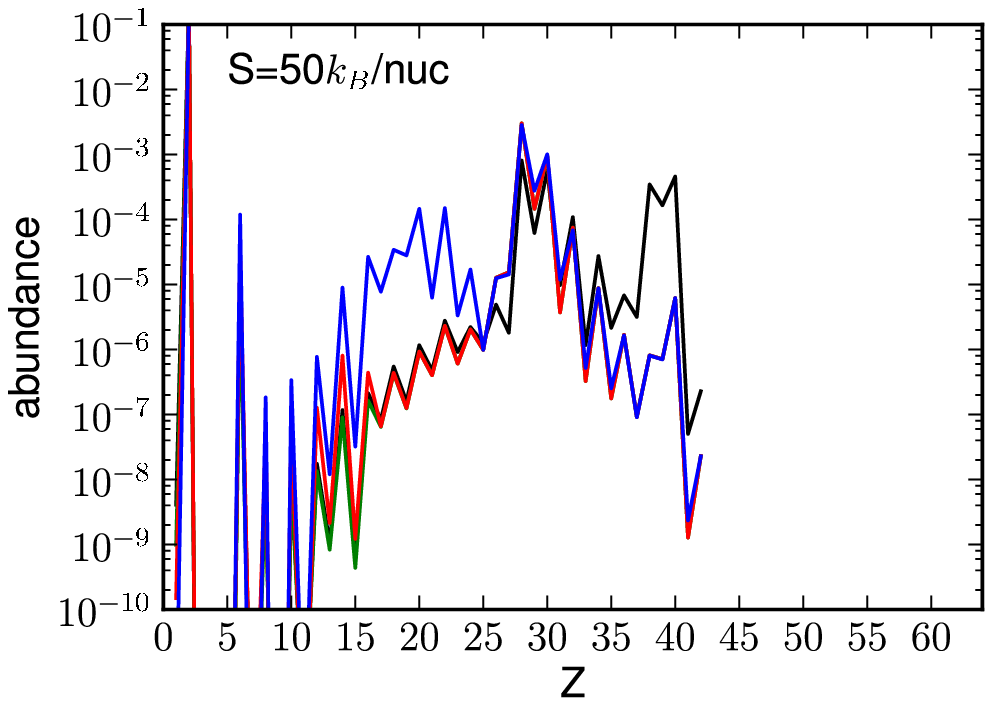}\\
  \includegraphics[width=8cm]{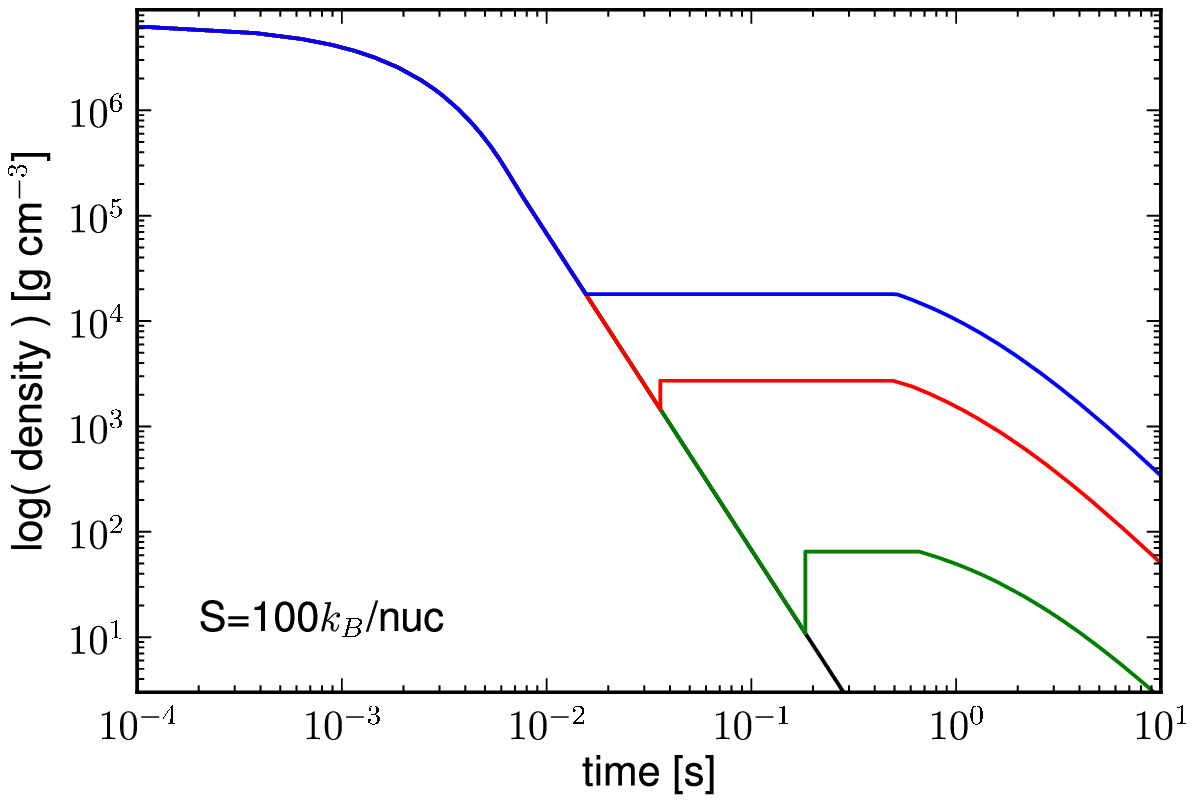}%
  \includegraphics[width=8cm]{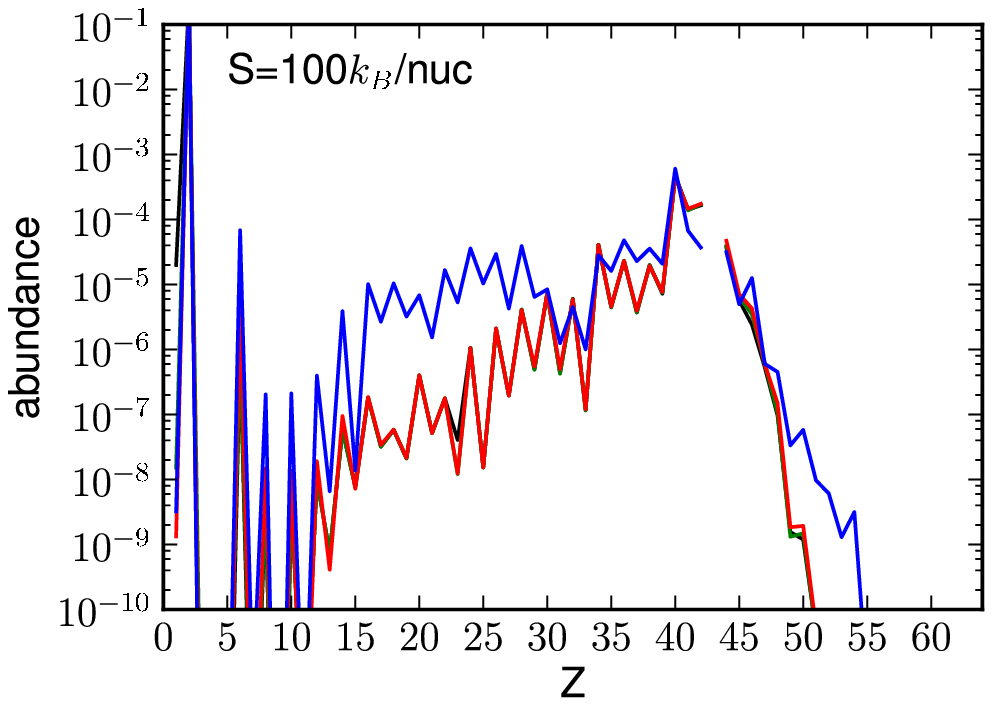}
  \caption{Left panels show the evolution of density assuming a wind
    termination at different positions. The expansion time scale is the
    same in both cases ($\tau=2$~ms) and the entropy is indicated in
    the figures. The resulting elemental abundances are shown in the
    right column.}
  \label{fig:rs_nrich}
\end{figure}

The final abundances also depend on the long-time evolution which can
be affected by the wind termination (see \ref{sec:par_rs} for a simple
parametrization). Figure~\ref{fig:rs_nrich} shows the density
evolution and resulting abundances for different positions of the wind
termination radius. Note that there is a shock, i.e. density jump,
only when the wind is supersonic.  The wind termination decelerates
the expansion, allowing for further production of nuclei up to the
iron group using alpha particles and neutrons.  In the case without
wind termination, the freeze-out of charged-particle reactions occurs
faster and the $Y_n/Y_{\mathrm{seed}}$ is higher. This allows to build
elements up to Sr, Y, Zr even when the entropy is low (upper
panel). When the entropy is high, the wind termination has some impact
on abundances only when it occurs at high temperatures. In this case,
charged-particle reactions are effective during longer time, leading
to higher abundances for $10<Z<40$. Independently of the wind
termination position, the neutron magic number $N=50$ is always
reached and matter flow stops there.

\begin{figure}[!htb]
  \includegraphics[width=8cm]{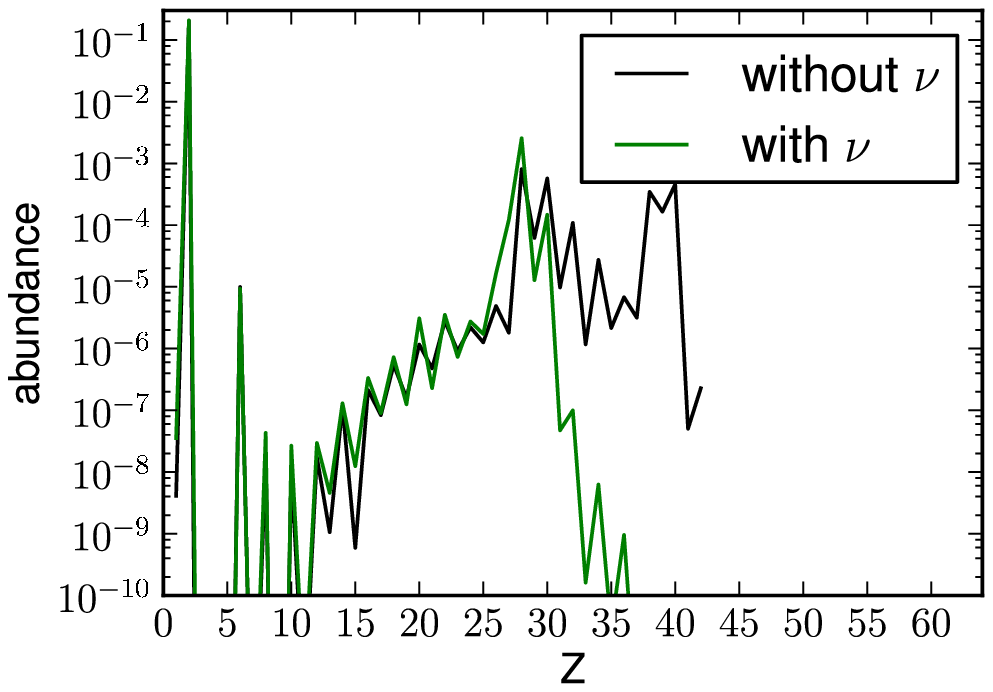}%
  \includegraphics[width=8cm]{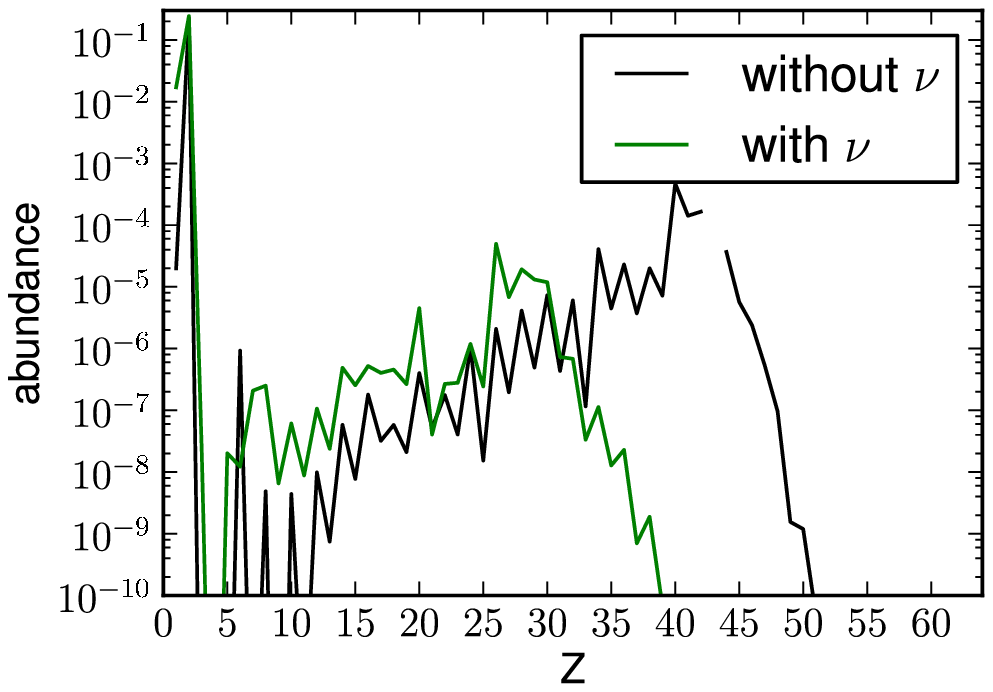}
  \caption{Elemental abundances with and without neutrinos for an
    expansion time scale of $\tau=2$~ms and $Y_e=0.49$. The entropy is
    $S=50,100 \, k_{\mathrm{B}}/\mathrm{nuc}$ in the left and right
    panels, respectively.}
  \label{fig:nrich_neut}
\end{figure}

Even more important than the reverse shock is the role of
neutrinos. When neutrino luminosities are high and the wind is
neutron-rich, electron neutrino absorption on neutrons lead to a
significant reduction of the neutron-to-seed ratio. The impact of
neutrinos on the weak r-process is presented in
Fig.~\ref{fig:nrich_neut}. When neutrinos are considered the
production of lighter heavy elements is strongly hindered. High
neutrino luminosities and energies may be found during the first
second after the explosion. At later times the destruction of neutrons
by neutrinos is less significant and lighter heavy elements are
synthesized.

\subsection{$\nu p$-process}
\label{sec:nup-process}
Hydrodynamical simulations of core-collapse supernovae with
sophisticated neutrino transport indicate that the early ejecta may be
proton rich instead of neutron-rich \cite{Liebendoerfer.etal:2003,
  Pruet.Woosley.ea:2005, Pruet.Hoffman.ea:2006,
  Froehlich.Hauser.ea:2006, Froehlich06, Buras.etal:2006a,
  Wanajo:2006}. Even long-time simulations suggest that spherically
symmetric neutrino-driven winds could stay proton rich during several
seconds \cite{Huedepohl.etal:2010, Fischer.etal:2010}. Although there
are still uncertainties in the determination of the $Y_e$, it is very
likely that the neutrino-driven wind turns proton rich during some
phase of the post explosion evolution.

In proton-rich condition elements beyond iron group can be synthesized
by the $\nu p$-process~\cite{Pruet.Hoffman.ea:2006, Froehlich06,
  Wanajo:2006}. Proton and alpha captures drive the matter flow up to
$^{64}$Ge which is a waiting point as its beta-decay half live is long
compared to the expansion time scale of the wind. However, the matter
is under a high neutrino flux and electron antineutrino absorption on
protons produces a small amount of neutrons. These extra neutrons lead
to $(n, p)$ reactions which are faster than beta decays and thus
permit the matter to flow towards heavier nuclei. The efficiency of
this process strongly depends on the expansion time scale, entropy,
electron fraction, but also on the neutrino luminosity and energies.

The most important role in the $\nu p$-process is played by neutrinos,
or more specifically by electron antineutrinos. These are responsible
to produce the necessary neutrons to overcome the bottlenecks. After
the initial very fast expansion of a wind trajectory, there is an
equilibrium between neutron capture and neutron production by
antineutrino absorption on protons. This implies that
\begin{equation}
  \frac{\mathrm{d}Y_n}{\mathrm{d}t} = \lambda_{\bar{\nu}_e}Y_p - \sum_{Z,A} n_n Y(Z,A) \langle \sigma v \rangle_{(Z,A)} =0 .
\end{equation}
Here $\lambda_{\bar{\nu}_e}$ is the electron antineutrino absorption
rate and $\langle \sigma v \rangle_{(Z,A)}$ is the sum of reaction
rates for $(n,\gamma)$ and $(n,p)$ reactions for nucleus
(Z,~A). Therefore, the neutron density in equilibrium is given by
\begin{equation}
  n_n = \frac{\lambda_{\bar{\nu}_e}Y_p }{\sum_{Z,A}Y(Z,A) \langle \sigma v \rangle_{(Z,A)}}.
\label{eq:n-equil}
\end{equation}
The electron antineutrino luminosities and energies strongly determine
the neutron density through $\lambda_{\bar{\nu}_e}$. Moreover, this
equation shows clearly the importance of the proton-to-seed ratio for
the $\nu p$-process.

\begin{figure}[!htb]
  \includegraphics[width=6cm]{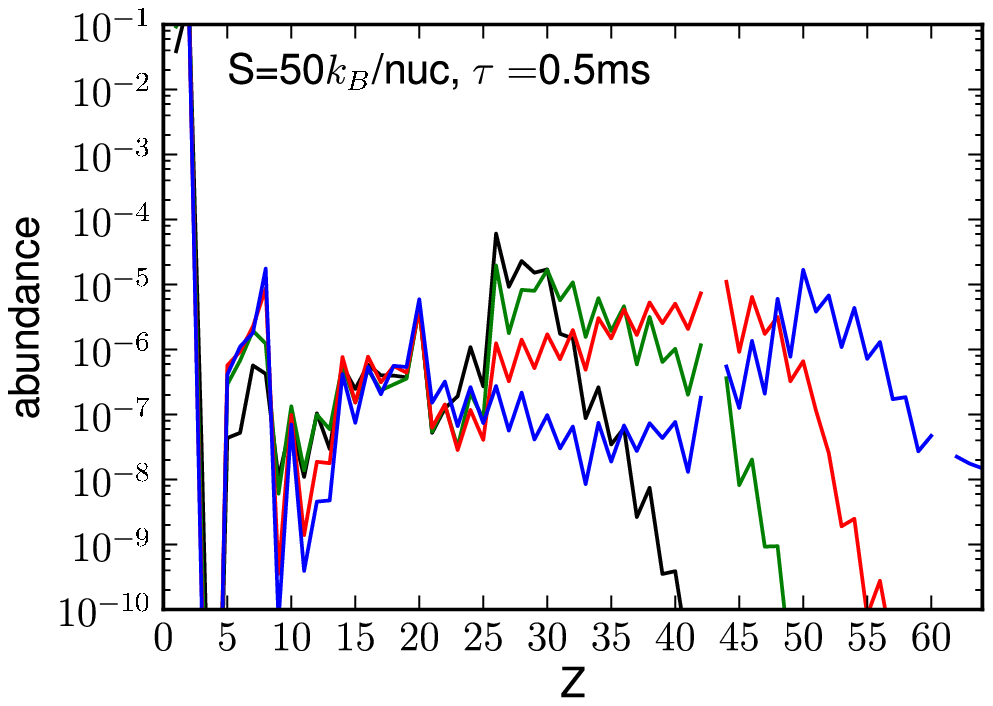}%
  \includegraphics[width=6cm]{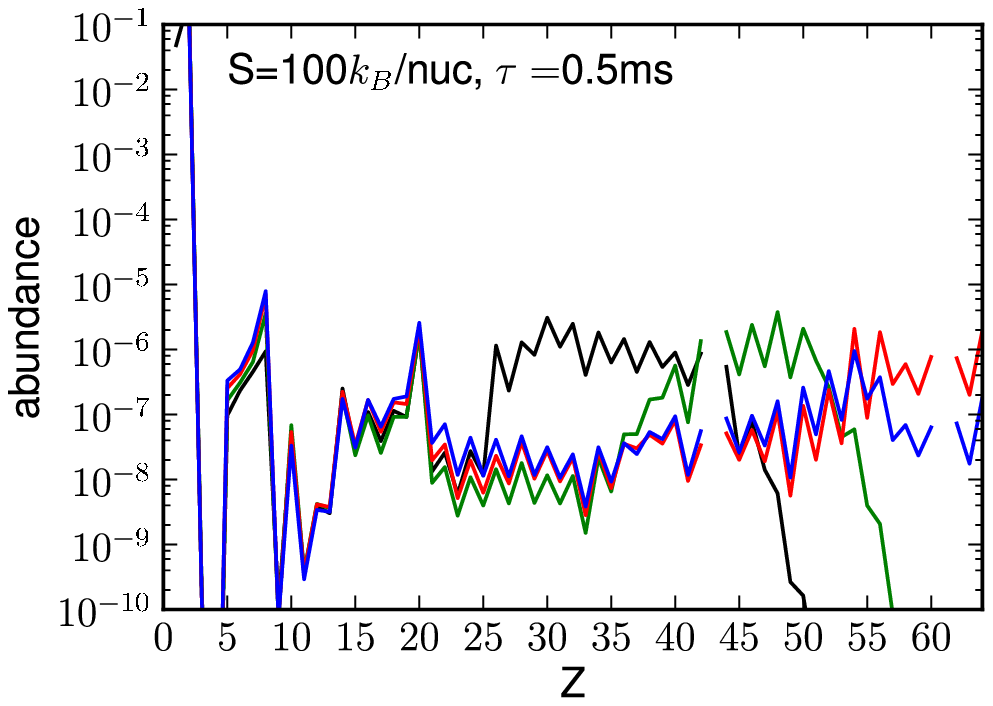}%
  \includegraphics[width=6cm]{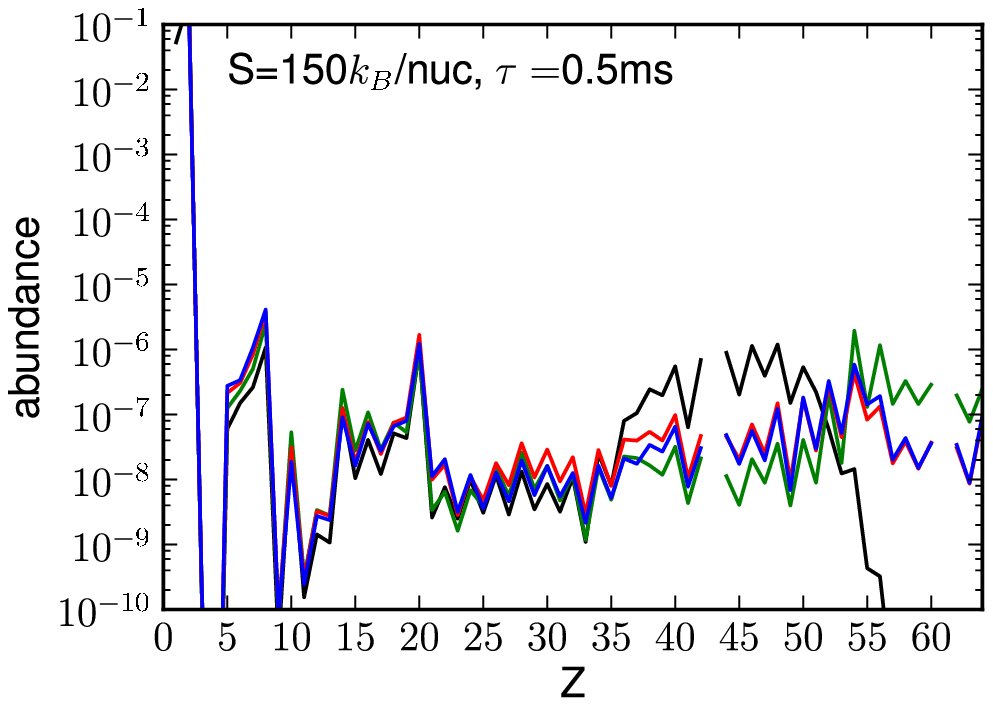}\\
  \includegraphics[width=6cm]{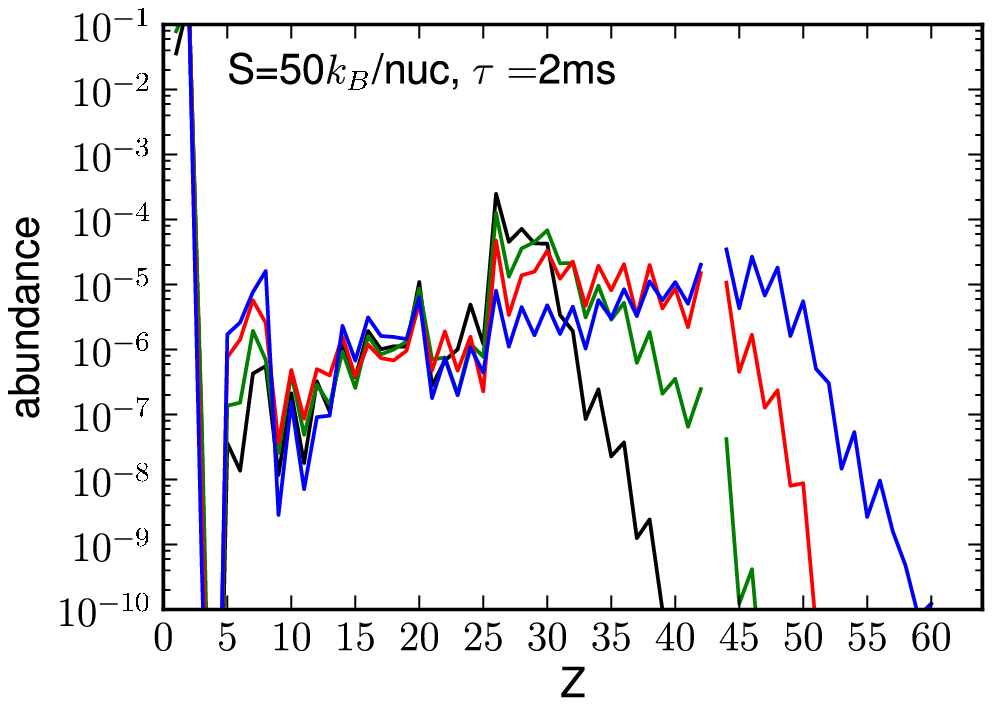}%
  \includegraphics[width=6cm]{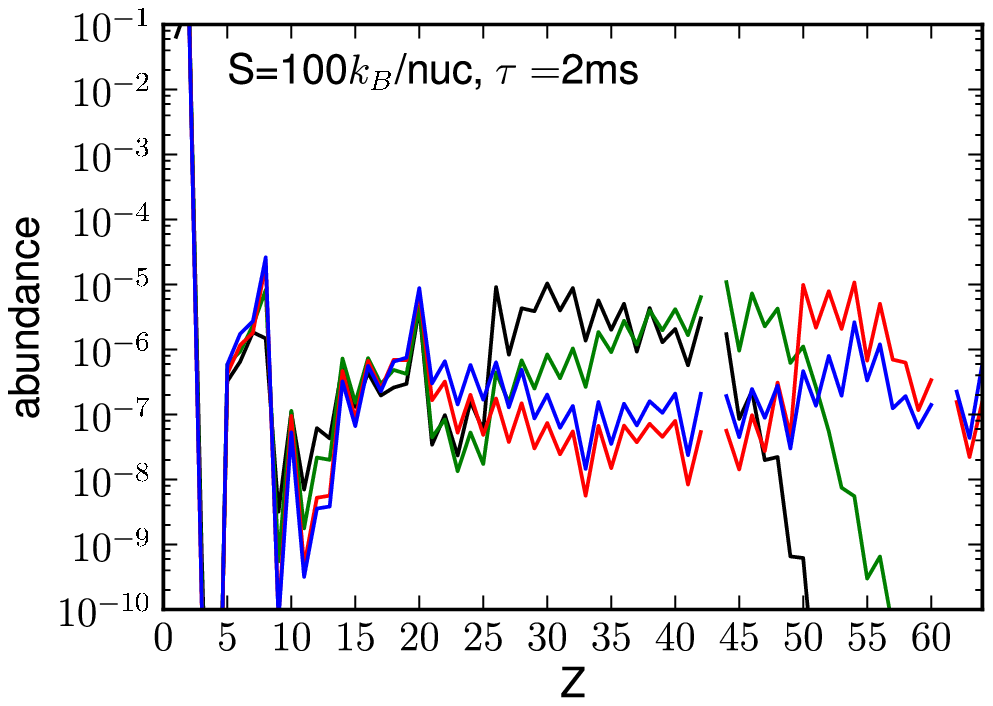}%
  \includegraphics[width=6cm]{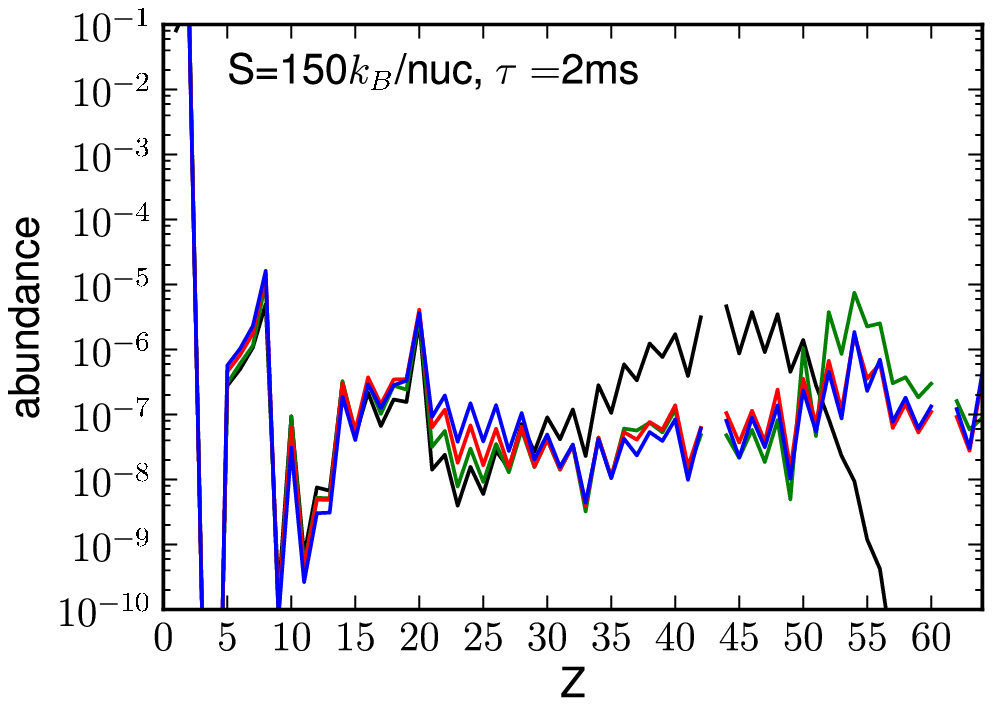}\\
  \includegraphics[width=6cm]{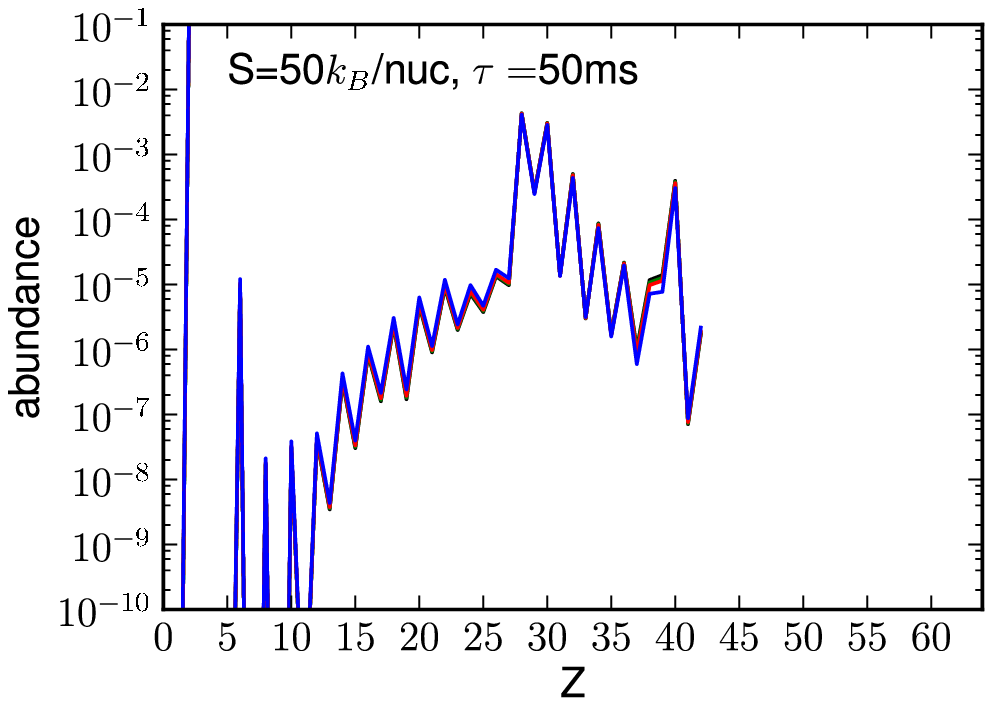}%
  \includegraphics[width=6cm]{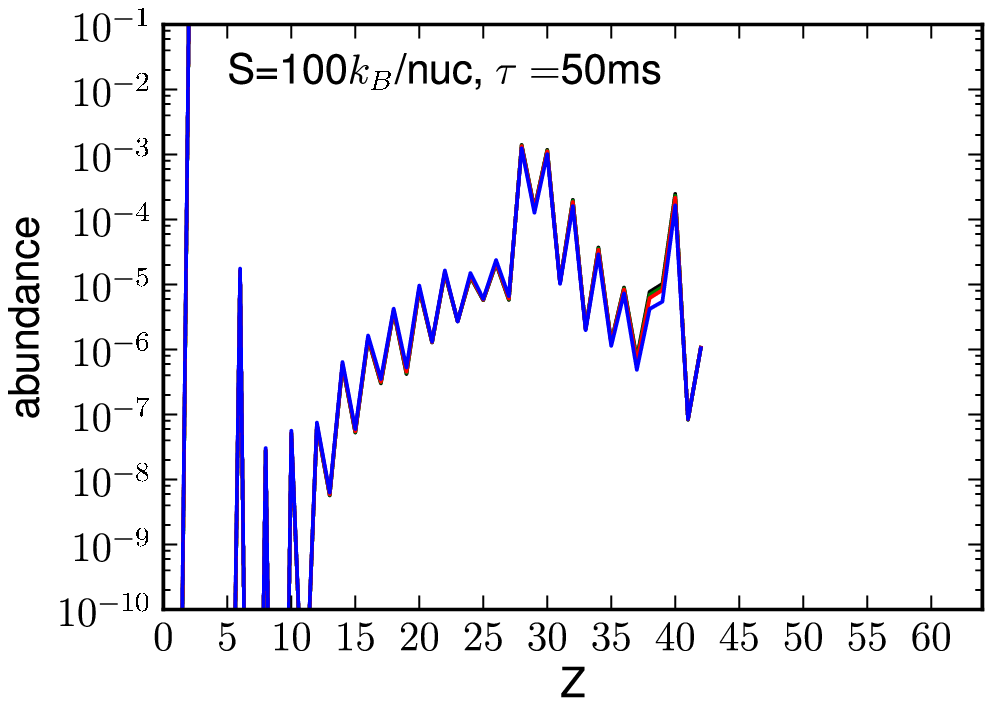}%
  \includegraphics[width=6cm]{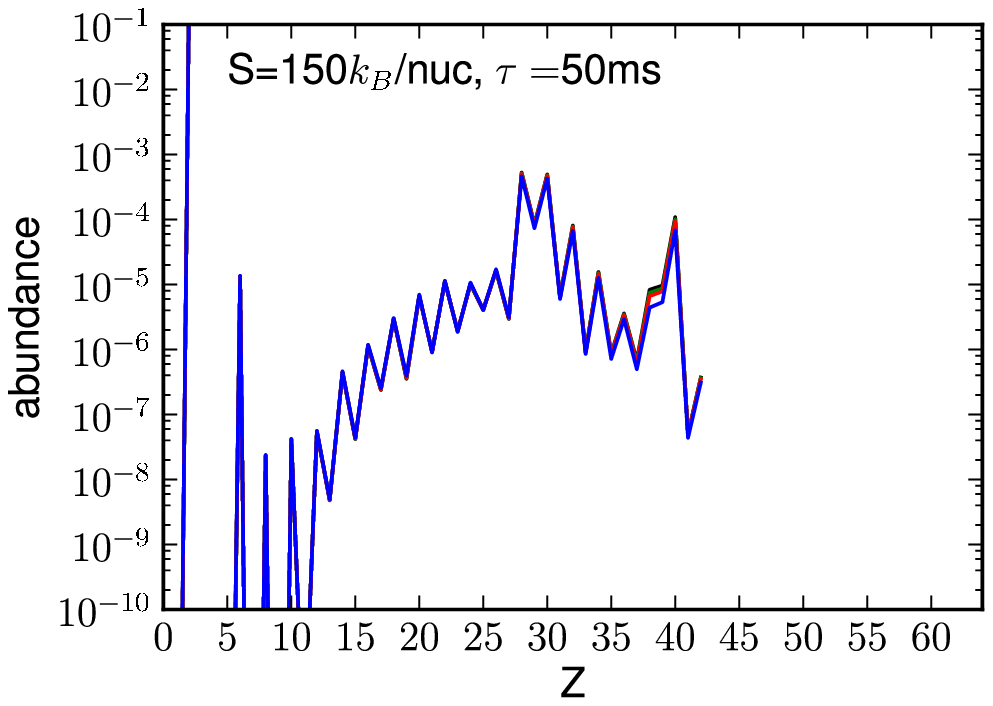}
  \caption{Elemental abundances for different electron fractions:
    $Y_e=$0.52 (black), 0.55 (green), 0.60 (red), 0.65 (blue). The
    entropy and expansion time scale are given in the figures.}
  \label{fig:YvsZ_prich}
\end{figure}

The dependence of the abundances on the wind parameters for
proton-rich condition is shown in Fig.~\ref{fig:YvsZ_prich}. Similar
to Fig.~\ref{fig:YvsZ_nrich} we present results for different
entropies in columns and various expansion time scales in rows. In
addition, every panel contains the abundances for three electron
fractions. The electron fraction affect the abundances as it
determines proton abundances which are key for the production of
neutrons as shown in Eq.~(\ref{eq:n-equil}).  The other two wind
parameters, i.e., entropy and expansion time scale, also influence the
proton-to-seed and neutron-to-seed ratios and therefore the final
abundances. As in neutron-rich conditions, high entropy prevents the
formation of seed nuclei and this results in higher
$Y_p/Y_{\mathrm{seed}}$ and $Y_n/Y_{\mathrm{seed}}$. The impact of the
expansion time scale follows also the same trend as in neutron-rich
winds. The initial build up of the seed nuclei is controlled by
three-body reactions which strongly depend on the expansion time scale
(see Sect.~\ref{sec:nuc_par}). This is extreme for the very slow
expansion, shown in the bottom panels of
Fig.~\ref{fig:YvsZ_prich}. Here all protons are consumed and final
abundances are dominated by alpha particles and nuclei up to $Z=40$.
There are two peaks in the abundances related to the region around the
waiting point $^{64}$Ge and to $A\approx 90$ where magic neutron
number $N=50$ and semi-magic proton number $Z=40$ are reached. For
faster expansions, the final abundances contain also protons because
the temperature drops too fast without leaving enough time for the
protons to be captured.

\begin{figure}[!htb]
  \includegraphics[width=8cm]{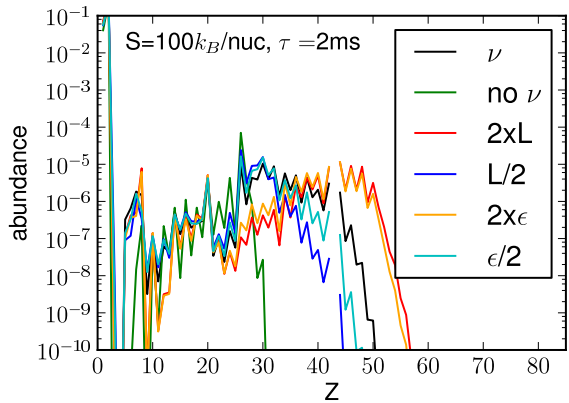}
  \caption{Elemental abundances with different neutrino treatment. The
    black line labeled as $\nu$ is the reference case and correspond
    to $Y_e=0.52$, $S=100 \, k_{\mathrm{B}} /\mathrm{nuc}$, and
    $\tau=2$~ms (middle panel of Fig.~\ref{fig:YvsZ_prich}). The other
    curves show calculations without neutrinos (no $\nu$), with
    neutrino luminosities increased (2$\times$L) or decreased (L/2) by
    a factor of two, and with same variations for the neutrino
    energies.}
  \label{fig:Y_prich_neut}
\end{figure}

In order to understand the sensitivity of the abundances to neutrino
luminosities and energies, we vary them for the wind with $Y_e=0.52$,
$S=100 \, k_{\mathrm{B}} /\mathrm{nuc}$, and $\tau=2$~ms. The
resulting abundances are shown in Fig.~\ref{fig:Y_prich_neut}. Without
neutrinos the matter stays mainly in $^{56}$Ni, which is the first
bottleneck before $^{64}$Ge. There is also a significant accumulation
of matter between $Z=26$ and $Z=32$ in the cases with reduced neutrino
luminosities and energies. However, the matter flow can continue
toward heavier nuclei with the small amount of neutrons produced.
When the neutrino luminosities or energies are increased by a factor
two, the $(n,p)$ reactions are more effective and heavier nuclei are
produced.

In contrast to the weak r-process, where the reverse shock has a minor
impact on the abundances, for the $\nu p$-process the wind termination
can significantly modify the abundances \cite{Wanajo.etal:2011,
  Arcones.Frohlich:2011}. The wind termination decelerates the
expansion allowing the matter to stay under the antineutrino flux for
longer time. Wanajo et al. (2011)~\cite{Wanajo.etal:2011} found that
there is an optimal temperature for the wind termination around 2~GK
to produce heavy elements. If the wind termination is much below this
temperature, the expansion is too fast and antineutrinos have not
sufficient time to produce the necessary neutrons to overcome
bottlenecks. This kind of very fast expansion is usually found in low
mass progenitors where there is not a termination shock and the $\nu
p$-process is very inefficient~\cite{Wanajo.Nomoto.ea:2009,
  Hoffman.Mueller.Janka:2008}. More massive progenitors lead to
smaller wind termination radius and thus higher
temperatures~\cite{arcones.janka.scheck:2007}. Very high temperature
at the wind termination ($T\approx3$~GK) hinder also the production of
heavy elements, as matter stays in the newly identified $^{59}$Cu
($p,\alpha$) $^{56}$Ni cycle~\cite{Arcones.Frohlich:2011}. When the
wind termination is around 2~GK the charged-particle reactions are
still very effective, a breakout from the NiCu-cycle can occur, and
antineutrino still produce neutrons leading also to efficient ($n,p$)
reactions. In addition to the temperature at the wind termination the
evolution afterwards is also key to determine the abundances. This
late evolution depends on the slow moving, early ejecta which are
strongly affected by multi-dimensional
anisotropies~\cite{Arcones.Janka:2011}. In some cases, the behaviour
of density and temperature after the wind termination may lead to
variations of the neutron density, allowing to move to stability not
only by $\beta$-decay but also by neutron
captures~\cite{Arcones.Janka:2011}. This is very important because it
implies that matter can cross to the neutron-rich side of stability
and even produce neutron-rich isotopes.

\section{Summary and outlook}
\label{sed:sum}

The present review discusses the mechanism of ejecting matter from the
(proto-)neutron star surface via energy deposition of the neutrinos
streaming out of the cooling and deleptonizing neutron star after a
successful supernova explosion (the so-called neutrino wind).  The
input physics entering this mechanism and its consequences for the
composition of the ejected material are described in detail.  The
latter is strongly dependent on the neutron star nuclear equation of
state, influencing the matter composition inside the neutron star, as
well as the neutrino interactions with matter, both determining the
final neutrino energy spectra and luminosities of all flavors.

While $\mu$ and $\tau$-neutrinos are dominated by neutral current
reactions, and pure scattering is not dominating the energy deposition
nor changing the composition, electron neutrinos and anti-neutrinos
act also via charged-current capture reactions (with strong energy
deposition) which also determine the overall proton-to-nucleon ratio
via $\nu_e + p \rightarrow n + e^+$ and $\bar\nu_e + n \rightarrow p +
e^-$. Due to the energy dependence (and Q-values) of the neutrino and
anti-neutrino cross sections, the composition becomes neutron-rich if
average energies of anti-neutrinos are larger than those of neutrinos
by 4$\times (m_n-m_p)c^2$, if both are characterized by similar total
luminosities. More generally also the luminosities of both species
enter.

Early calculations in the 90s resulted in substantial mean energy
differences between anti-neutrinos and neutrinos, caused a
neutron-rich composition and high entropies of the ejected matter. The
entropy $S$, total proton-to-nucleon ratio $Y_e$, and the expansion
time scale $\tau$ of ejected matter are the key properties for the
composition of the ejecta (possibly still modified by late time
behavior not corresponding to free expansions if matter experiences
reverse shocks due to collision with matter ejected in the prior
supernova explosion).  These early calculations predicted a large
neutron-to-seed ratio after freeze-out of charged particles reactions,
providing the basis for a strong r-process and the production of heavy
nuclei up to Th and U. This made the neutrino wind the most promising
site for the r-process nuclei found in the solar composition and
indicated a very strong production of r-process matter in the early
Galaxy, as massive stars producing supernovae are the fastest evolving
species and the earliest polluters of the interstellar medium
(consistent with observations of low metallicity stars). These
investigations also caused a large amount of parametrized
calculations, studying r-process properties as a function of the three
key parameters $S$, $Y_e$, and $\tau$.

More recent core collapse simulations with late-time evolutions up to
10~s after core collapse, and improved micro physics and neutrino
transport, resulted in anti-neutrino and neutrino spectra with smaller
mean energy differences, and consequently a proton-rich composition of
ejecta. This gave rise to a new process, the $\nu$p-process, where in
a first step the more proton-rich conditions produce essential
abundances of $^{64}$Ge (decaying into $^{64}$Zn, thus going beyond
$^{56}$Ni decaying into $^{56}$Fe) which could explain that Zn is
co-produced with Fe-group nuclei, as observed in low-metallicity
stars. In a second step portions of $^{64}$Ge can be processed further
up to A=90 nuclei, due to overcoming its long beta-decay half-life via
an $(n,p)$-reaction with neutrons, produced by anti-neutrino captures
on the remaining free protons in this proton-rich environment. This
opened an option to explain abundances of light p-nuclei.

Recent calculations, including medium effects for neutrons and
protons, which lead to effective Q-values corrected by chemical
potential differences have apparently again provided a change. The
full consequences of this effect are not fully analyzed, yet, and
surprises are still expected. In any case, both proton-rich and
slightly neutron-rich conditions are adequate to synthesized lighter
heavy elements such as Sr, Y, Zr. Their abundances have been
attributed before to a not-yet understood lighter (heavy) element
primary process (LEPP) which is required in galactic evolution based
on low-metallicity observations.

In addition, the final understanding will probably only arise with the
full understanding of core collapse supernova explosions from 3D
hydrodynamical modeling. The latter is well on its way, but not yet
established.  Thus, the neutrino wind properties will probably still
provide further surprises and possibly even contain matter with
properties covering all cases discussed above, due to the time
evolution of neutrino spectra and luminosities.

\ack 

A.A. was supported by a Feodor Lynen Fellowship (Humboldt Foundation)
and by the Helmholtz-University Young Investigator grant
No. VH-NG-825. F.K.T. acknowledge support by the Humboldt Research
Award. We acknowledge support and funding by the Swiss National Science
Foundation (SNF).  The authors are additionally supported by
EuroGENESIS, a collaborative research program of the European Science
Foundation (ESF) and the Helmholtz Nuclear Astrophysics Virtual
Institute (NAVI, VH-VI-417).

\appendix

\section{Parametric wind trajectories}
\label{sec:par_traj}

We give an useful parametrization for wind trajectories based on
hydrodynamical simulations. The evolution of density is assumed to be
first exponential and then to follow a power law. The wind entropy is
constant as the expansion is adiabatic. The reverse shock leads to an
increase of density and entropy that will be also parametrized in the
next section. Once density and entropy are known the temperature is
obtained from an equation of state (see e.g., \cite{Witti.etal:1994,
  Freiburghaus.Rembges.ea:1999}). The evolution of radius is provided
by mass conservation in a steady state outflow: $\dot{M}=4\pi r^2 v
\rho = $~constant with $v=\mathrm{d}r / \mathrm{d}t$.

Here we assume the density drops exponentially between $T=10$~GK and a
lower limit temperature $T_{\mathrm{exp}}$. This can be used as an
additional parameter, although here we fix it to
$T_{\mathrm{exp}}=4$~GK, i.e., the expansion is exponential while
charged particle reactions dominate the evolution. Previous works
assumed also a mixed evolution from exponential to power law (e.g.,
Ref.~\cite{Meyer:2002}) and they use an additional time scale to
switch from one to the other.

For $T>T_{\mathrm{exp}}$, density, radius, and velocity at a time
$t>t_0$ (with $t_0$ being the initial time which fulfils
$T(t_0)\approx 10$~GK) are given by:
\begin{eqnarray}
  \rho(t) &=& \rho_0 e^{(t_0-t)/\tau} \label{eq:par_den_exp}\\
  r(t) &=& r_0 \left[1 + 3\frac{v_0}{r_0} \tau 
    \left( e^{(t_0-t)/\tau} - 1 \right) \right]^{1/3} \label{eq:par_rad_exp}\\
  v(t) &=& v_0 e^{(t_0-t)/\tau}
  \left[1 + 3\frac{v_0}{r_0} \tau \left( e^{(t_0-t)/\tau} - 1 \right) \right]^{-2/3} \label{eq:par_vel_exp}
\end{eqnarray}
Here the quantities $\rho_0$, $r_0$, and $v_0$ correspond to density,
radius, and velocity at the initial time $t_0$, respectively. The
expansion time scale is $\tau$.

After the temperature drops below $T\sim T_{\mathrm{exp}}$ the
evolution continues as a power law:
\begin{eqnarray}
  \rho(t) &=& \rho_{\mathrm{exp}} \left(\frac{t_{\mathrm{exp}}}{t} \right)^3 \label{eq:par_den_pl}\\
  r(t)    &=& r_{\mathrm{exp}} \left[ 
                        1 + \frac{3}{4} \frac{v_{\mathrm{exp}} t_{\mathrm{exp}}}{r_{\mathrm{exp}}} 
                        \left( \left(\frac{t}{t_{\mathrm{exp}}}\right)^4 -1 \right)  
                  \right]^{1/3}  \label{eq:par_rad_pl}\\
  v(t)    &=& v_{\mathrm{exp}} \left(\frac{t}{t_{\mathrm{exp}}}\right)^3
                  \left[ 
                        1 + \frac{3}{4} \frac{v_{\mathrm{exp}} t_{\mathrm{exp}}}{r_{\mathrm{exp}}} 
                        \left( \left(\frac{t}{t_{\mathrm{exp}}}\right)^4 -1 \right)  
                  \right]^{-2/3}  \label{eq:par_vel_pl}
\end{eqnarray}
The quantities $\rho_{\mathrm{exp}}$, $r_{\mathrm{exp}}$,
$v_{\mathrm{exp}}$, and $t_{\mathrm{exp}}$ correspond to the moment
when the temperature is $T=T_{\mathrm{exp}}$. Note that the velocity
increases for $t \rightarrow \infty$, although in neutrino-driven
winds the velocity asymptotically converges to a constant value. This
can be cured by using a power law with $\rho \propto t^{-2}$ (see
e.g., Ref.\cite{Panov.Janka:2009,Wanajo.etal:2011} or by including the
deceleration due to the interaction with the slow moving ejecta
(see~\ref{sec:par_rs}).

\section{Reverse shock}
\label{sec:par_rs}
The main impact of the reverse shock on the evolution is the sudden
deceleration of matter, which leads a drop of the expansion velocity
and an increase of temperature and density, as kinetic energy is
transformed into internal energy.  This increase can be determined
using the Rankine-Hugoniot conditions to calculate the values of the
density ($\rho_{\mathrm{rs}}$), temperature ($T_{\mathrm{rs}}$), and
velocity ($u_{\mathrm{rs}}$) of the shocked material relative to the
wind values (marked by the subscript ``$w$''). The Rankine-Hugoniot
conditions for mass, momentum, and energy conservation through the
shock and are given by:
\begin{eqnarray}
  \label{eq:rh}
  \rho_{\mathrm{w}}u_{\mathrm{w}} &=
  \rho_{\mathrm{rs}}u_{\mathrm{rs}} \label{eq:rh1}\\
  P_{\mathrm{w}} + \rho_{\mathrm{w}}u_{\mathrm{w}}^2 & =
  P_{\mathrm{rs}} + \rho_{\mathrm{rs}}u_{\mathrm{rs}}^2  \label{eq:rh2}\\
  \frac{1}{2}u_{\mathrm{w}}^2 + \epsilon_{\mathrm{w}}+
  \frac{P_{\mathrm{w}}}{\rho_{\mathrm{w}}} &= \frac{1}{2}u_{\mathrm{rs}}^2 +
  \epsilon_{\mathrm{rs}}+
  \frac{P_{\mathrm{rs}}}{\rho_{\mathrm{rs}}} \label{eq:rh3}
\end{eqnarray}
here $\rho$, $v$, $P$, and $\epsilon$ are the density, velocity,
pressure, and specific internal energy, respectively. In the last
equation one can take $\epsilon \approx 3P/\rho$ because it is
radiation dominated environment. Assuming that the lhs of these
equations is known, thus combining the three equations one gets two
possible solutions for the matter velocity after the shock: 1)
$u_{\mathrm{rs}} = u_{\mathrm{w}}$, no shock; 2) $u_{\mathrm{rs}}=
u_{\mathrm{w}}/7 + 8/7 \,
P_{\mathrm{w}}/(\rho_{\mathrm{w}}u_{\mathrm{w}})$. Once the velocity is
known, the density and pressure can be computed using
Eqs.~(\ref{eq:rh1}),~(\ref{eq:rh2}). 

The evolution of density after the reverse shock can be determined
from the condition of constant mass outflow: $\dot{M} = 4\pi r^2 v
\rho$. There are two extreme possibilities: 1) velocity is constant
and density decreases as $r^{-2}$; 2) density is constant and
therefore velocity drops as $r^{-2}$. The latter was used in
Ref.~\cite{Kuroda.Wanajo.Nomoto:2008}. However, it implies a decrease
of the velocity down to a few m~s$^{-1}$ in about a second, that is
not consistent with simulations ~\cite{arcones.janka.scheck:2007,
  Fischer.etal:2010}, where the post-shock velocities are around
$10^3$--$10^4$~km~s$^{-1}$. The simulations suggest something in
between these two extremes. First the density stays almost constant
during $\approx 0.5-1$~s and the velocity decreases as $r^{-2}$, later
the velocity stays constant and consequently the density decreases as
$r^{-2}$. Once the density is known the temperature can be determined
from the condition of constant entropy (adiabatic expansion).

\begin{figure}[!htb]
  \includegraphics[width=8cm]{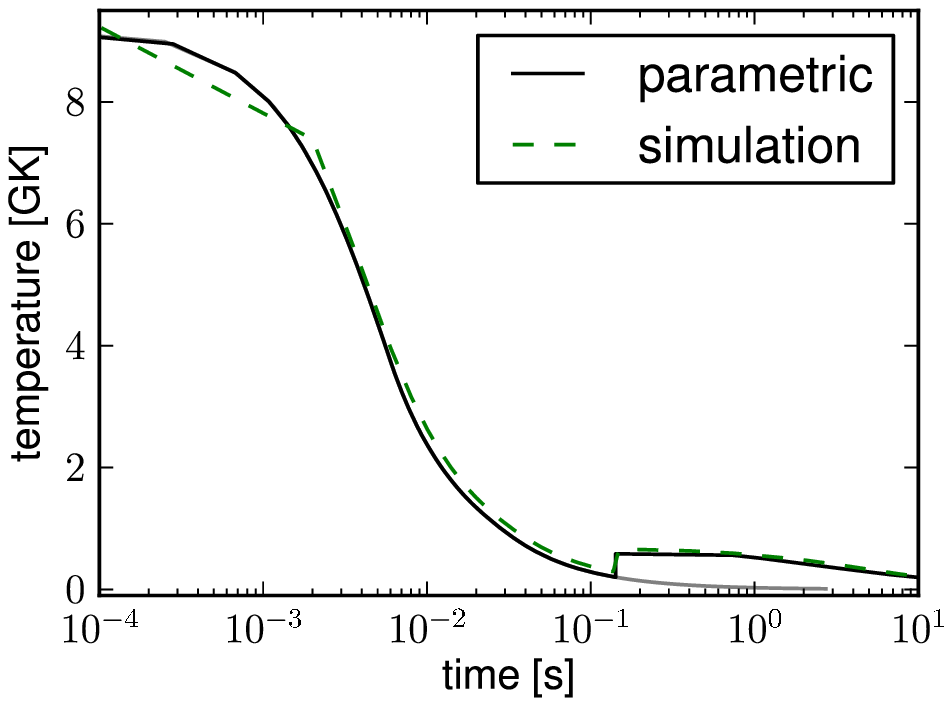}
  \includegraphics[width=8cm]{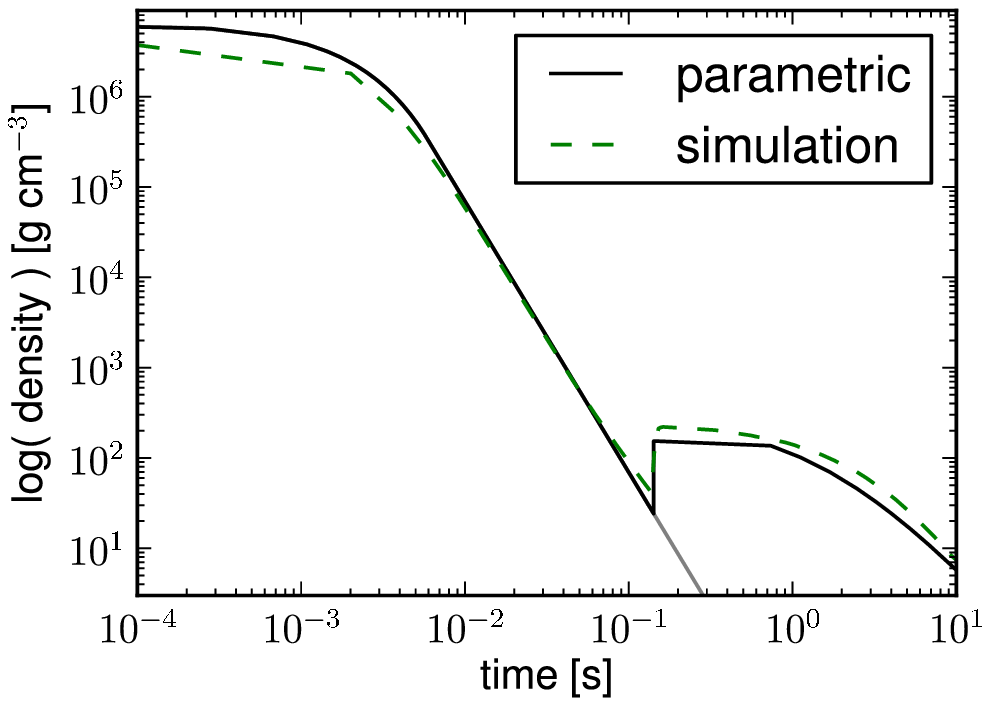}
  \caption{Comparison of temperature and density evolutions from the
    supernova simulation (trajectory ejected 5~s after bounce of model
    M15l1r1 of Ref.~\cite{arcones.janka.scheck:2007}) to the evolution
    based on our parametric description. The grey solid line
    corresponds to the evolution without wind termination
    (~\ref{sec:par_traj}).}
  \label{fig:par_Td_rs}
\end{figure}

A comparison between the parametric expansion and the results from the
simulations is presented in Figure~\ref{fig:par_Td_rs}. The grey line
corresponds to the expanion without reverse shock as introduced in
\ref{sec:par_traj}. The dashed green line from the simulation has
lower resolution than the parametrization. This explains the initial
differences.

\section*{References}

\end{document}